\documentclass[prb,twocolumn,nofootinbib, showpacs]{revtex4}

\usepackage{amssymb, amsfonts, amsmath,dsfont,epsf,pstricks,graphicx,epsfig,epic,eepic,psfrag,latexsym, bm, verbatim, rotating,multirow,wasysym,marvosym,feyn}

\newcommand{\qq}{\mathbf{q}}

\newcommand{\dd}{\mathbf{d}}
\newcommand{\pp}{\mathbf{p}}
\newcommand{\nn}{\mathbf{n}}
\newcommand{\rr}{\mathbf{r}}
\newcommand{\hh}{\mathbf{h}}

\newcommand{\QQ}{\mathbf{Q}}
\newcommand{\dsum}{\displaystyle\sum}
\newcommand{\dprod}{\displaystyle\prod}
\newcommand{\dint}{\displaystyle\int}
\newcommand{\doint}{\displaystyle\oint}
\newcommand{\n }{{\cal N}}

\graphicspath{{./}{Figs/}{Plots/}}

\psfrag{``Pi''}{\Large{$\Pi$}}
\psfrag{``pi''}{\Huge{$\pi$}}
\psfrag{``G''}{\Large$G$}
\psfrag{``seq''}{\huge$\simeq$}
\psfrag{``equiv''}{\Large$\equiv$}
\psfrag{``rz''}{$r_{z}$}
\psfrag{``rx''}{$r_{x}$}
\psfrag{``ry''}{$r_{y}$}
\psfrag{``hz''}{\textcolor{red}{$h_{z}$}}
\psfrag{``hx''}{\textcolor{red}{$h_{x}$}}
\psfrag{``hy''}{\textcolor{red}{$h_{y}$}}
\psfrag{``r1''}{$r_{3}$}
\psfrag{``r2''}{$r_{2}$}
\psfrag{``r3''}{$r_{1}$}
\psfrag{``h1z''}{$h_{z}\sigma^{z}_{3}$}
\psfrag{``h2y''}{$h_{y}\sigma^{y}_{2}$}
\psfrag{``h3x''}{$h_{x}\sigma^{x}_{1}$}
\psfrag{``x''}{\textcolor{blue}{x}}
\psfrag{``y''}{\textcolor{magenta}{y}}
\psfrag{``z''}{\textcolor{red}{z}}
\psfrag{``pos1''}{1}
\psfrag{``pos2''}{2}
\psfrag{``pos3''}{3}
\psfrag{``pos4''}{4}
\psfrag{``pos5''}{5}
\psfrag{``pos6''}{6}
\psfrag{``plaq''}{P}
\psfrag{``r1A''}{$r_{1}$}
\psfrag{``rvA''}{$r_{v1}$}
\psfrag{``r1B''}{$r_{2}$}
\psfrag{``rvB''}{$r_{v2}$}

\begin{document}
\title{Site dilution in Kitaev's honeycomb model}
\author{A.\ J.\ Willans$^{1}$}
\author{J.\ T.\ Chalker$^{1}$}
\author{R.\ Moessner$^{2}$}
\date{\today}
\affiliation{$^{1}$Theoretical Physics, Oxford University, 1, Keble Road,  Oxford OX1 3NP, United Kingdom\\$^{2}$Max-Planck-Institut f\"{u}r Physik komplexer Systeme, N\"othnitzer Stra{\ss}e 38, 01187 Dresden, Germany}

\begin{abstract}
We study the physical consequences of site dilution in Kitaev's honeycomb model, in both its gapped and gapless phases. We show that a vacancy binds a flux of the emergent $Z_2$ gauge field and induces a local moment. In the gapped phase this moment is free while in the gapless phase the susceptibility has the dependence $\chi(h)\sim\ln(1/h)$ on field strength $h$. Vacancy moments have interactions that depend on their separation, their relative sublattice, and the phase of the model. Strikingly, in the gapless phase, two nearby vacancies on the same sublattice have a parametrically larger $\chi(h)\sim(h[\ln(1/h)]^{3/2})^{-1}$. In the gapped phase, even a finite density of randomly distributed vacancies remains tractable, via a mapping to a bipartite random hopping problem. This leads to a strong disorder form of the low-energy thermodynamics, with 
a Dyson-type singularity in the density of states for excitations.
\end{abstract}

\maketitle

\section{Introduction}
The interplay of disorder and interactions is one of the most fascinating aspects of condensed matter physics. It is also one of the most challenging ones, and opportunities for an exact analysis are rare.
We show in this paper that Kitaev's honeycomb model \cite{Kitaev:2006ys} offers a new and fruitful setting for such investigations. 

The introduction of randomness places the Kitaev model in the broader context of spin systems with quenched disorder, where the physics of spin glasses and infinite randomness phases are just two instance of conceptually new physics which have arisen.
More specifically, there is considerable interest in the question what happens when quenched disorder is introduced into a magnetic system exhibiting a novel correlated ground state, such as a classical or quantum spin liquid. Besides a search for new types of disorder-induced phases, it turns out that the properties of many quantum systems can be probed sensitively through the controlled introduction of impurities. In particular, impurities may reveal elusive features of the clean system, as illustrated by the use of non-magnetic ions to uncover the order-parameter symmetry in a superconductor \cite{Alloul:2009}. 
Experimentally, local probes such as nuclear magnetic resonance or atomic force microscopy can be used to distinguish impurity from bulk susceptibilities \cite{Alloul:2009}.

An example of this behaviour in one dimension is site dilution of an antiferromagnetic spin-$\frac{1}{2}$ Heisenberg chain, which creates free chain ends \cite{Eggert:1992}, leading to a Curie contribution to the susceptibility \cite{Sirker:2007}. In higher dimensions, putative examples of spin liquids have in particular attracted attention, as for these no local diagnostic, such as an order parameter, is available for determining their nature. For instance, numerical work on perhaps the most enigmatic $S = 1/2$ Heisenberg magnet in $d = 2$, that on the kagome lattice, indicates that non-magnetic impurities generate a local dimerisation pattern but do not induce a local moment \cite{Dommange2003}. In an opposite extreme, the classical (gapless) spin liquid on the SCGO lattice exhibits a local moment evidencing classical fractionalisation: in the low-temperature limit, its size is exactly half that of a free spin's \cite{Sen:2011}. This goes along with an extended spin texture visible in a modulated local susceptibility in NMR experiments,\cite{Limot:2002} a feature also predicted to occur for a candidate gapless quantum spin liquids on the kagome lattice \cite{Gregor:2009}. 

On the level of a theoretical description, we are limited by the small number of instances of spin Hamiltonians for which a simple and controlled derivation of the existence and nature of a quantum spin liquid phase is available. Happily, the Kitaev honeycomb model provides a rare example of a solvable spin model with both gapped and gapless liquid phases \cite{Kitaev:2006ys}. As a particular attraction, solvability in this context implies not only detailed knowledge of the respective ground states, but also the availability of much information on the excited states.

The Kitaev honeycomb model has attracted much interest in its own right as it exhibits some of the necessary elements to  develop a quantum computer: it supports fractionalized excitations, both Abelian and non-Abelian. There have been several proposals for an experimental realisation of the model, both with cold atoms in an optical lattice \cite{Demler:2003} and in a solid state system \cite{Jackeli:2009, Jackeli:2010}.

The solvability of the model has its origin in the existence of an extensive set of non-dynamical fluxes of an emergent Z$_{2}$ gauge field that permits a reduction of the Hamiltonian to a free fermion problem. We show in this paper that these steps remain tractable in the presence of site dilution and the Hamiltonian remains solvable, allowing us to calculate the magnetic properties of the Kitaev honeycomb model in the presence of vacancies. 

As one central result, we find that a vacancy binds a Z$_{2}$  flux. Since these fluxes are the aforementioned anyons \cite{Kitaev:2006ys}, this may have consequences for quantum computation: computations are performed by braiding the fluxes and if those braids encircle an impurity an additional Aharonov-Bohm phase may be picked up.

Beyond this, gapped and gapless phases differ greatly in their response to the introduction of non-magnetic impurities. In the gapped phase, we find that a single vacancy generates a paramagnetic moment with a magnitude that tends to that of a free spin as the gap becomes large. It is localised on one site adjacent to the vacancy. In the gapless phase, interactions with bulk excitations lead to an effective field-dependent moment, with a divergent susceptibility $\chi(h)\sim \ln(1/h)$, localised now on all three sites adjacent to the vacancy. 

The moments of different vacancies interact, in ways that depend on the phase of the system and the relative sublattice of the vacancies. In the gapless phase, the most dramatic consequence arises for two nearby vacancies on the same sublattice. Here the impurity susceptibility is parametrically larger than for a single vacancy $\chi(h)\sim(h[\ln(1/h)]^{3/2})^{-1}$. In the gapped phase, the interaction between vacancies on different sublattices decreases exponentially with their separation. 
In a system with many vacancies
we obtain a situation akin to the picture of the Bhatt-Lee singlet phase. This 
it turns out can be analysed
as a random bipartite hopping problem. We characterise the resulting broad distributions of energy levels underpinning the thermodynamics, with a density of states $\rho (E) \sim \mathcal{F} (E)/E$, where $\mathcal{F} (E)$ vanishes slower than any power of energy $E$ as $E \to 0$.

In the remainder of this paper, we flesh out these assertions. We begin by introducing the model and its emergent degrees of freedom along with useful notation in Sec.~\ref{TheModel}. Sec.~\ref{ModelingVacancies} presents the low-energy Hamiltonian in presence of vacancies and magnetic field. The gapped phase is discussed in Sec.~\ref{GappedPhase}. A perturbative demonstration of flux-binding is followed by successive treatment of the properties of the one-, two-, and many-vacancy problem. Sec.~\ref{GaplessPhase} is devoted to the gapless phase. Here we provide details for the analysis of the behaviour of an isolated vacancy and a vacancy pair. Sec.~\ref{Summary} summarises with a brief discussion and pointers to open questions. Two appendices treat some more technical material: how to project from the enlarged Majorana fermion Hilbert space down to the physical $S = 1/2$ Hilbert space;  and details of the honeycomb lattice Green functions, which we use extensively.

A short account of some of this work, particularly that on the gapless phase, has appeared in a Letter \cite{willans2010}, which also covered weak bond disorder. Some details of calculations omitted from the present paper are described in Ref.\onlinecite{Willans_thesis}.  Before proceeding, we alert the reader to a superficially different but conceptually related study of the coupling of a magnetic impurity to a spin in the Kitaev honeycomb model\cite{Tripathi:2009}.

\begin{figure}[h]
\begin{minipage}[b]{0.5\linewidth}
\epsfig{file=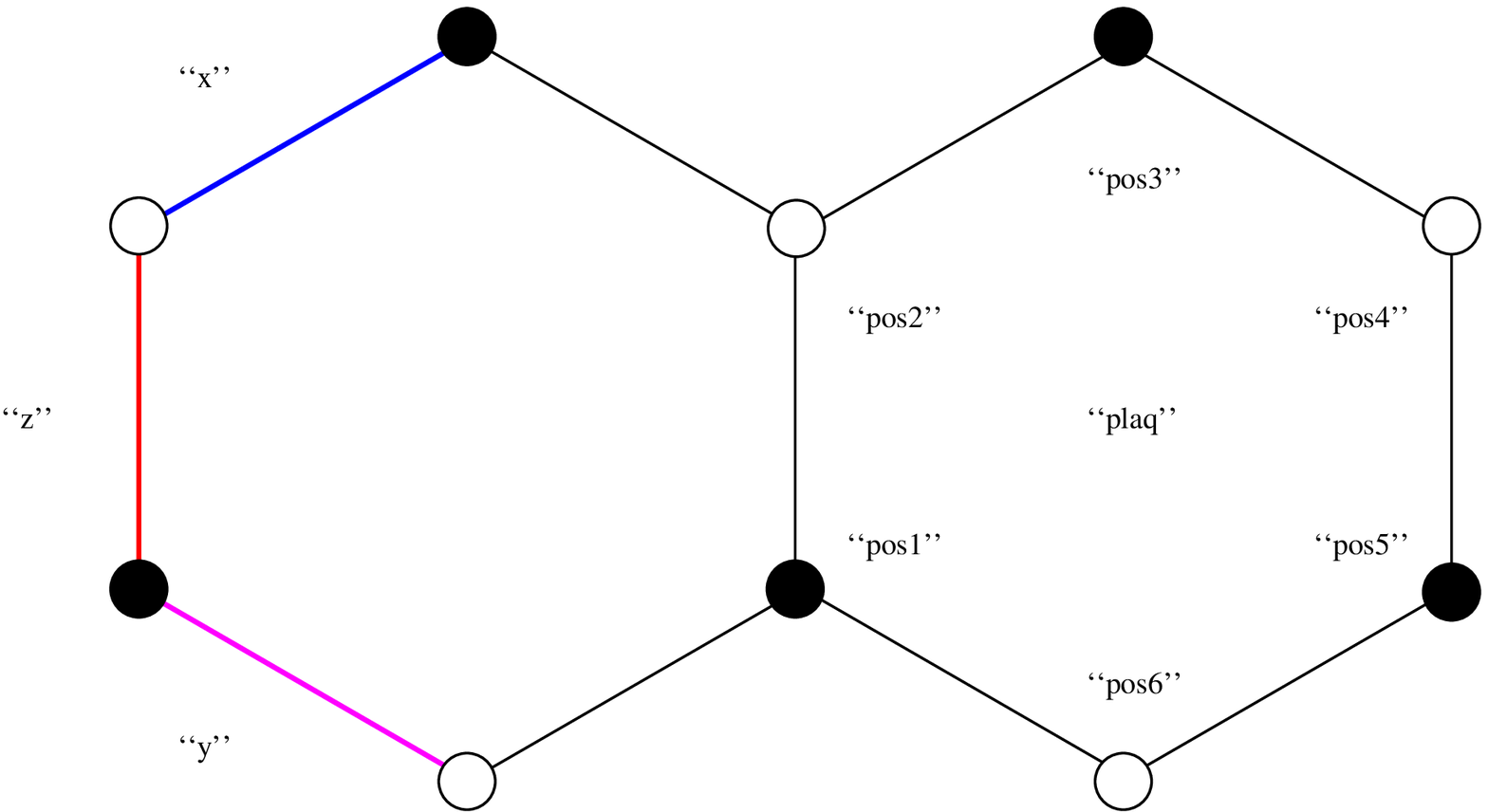, width=3.5cm}
\vspace{0.1\linewidth}
\end{minipage}
\begin{minipage}[b]{0.35\linewidth}
\epsfig{file=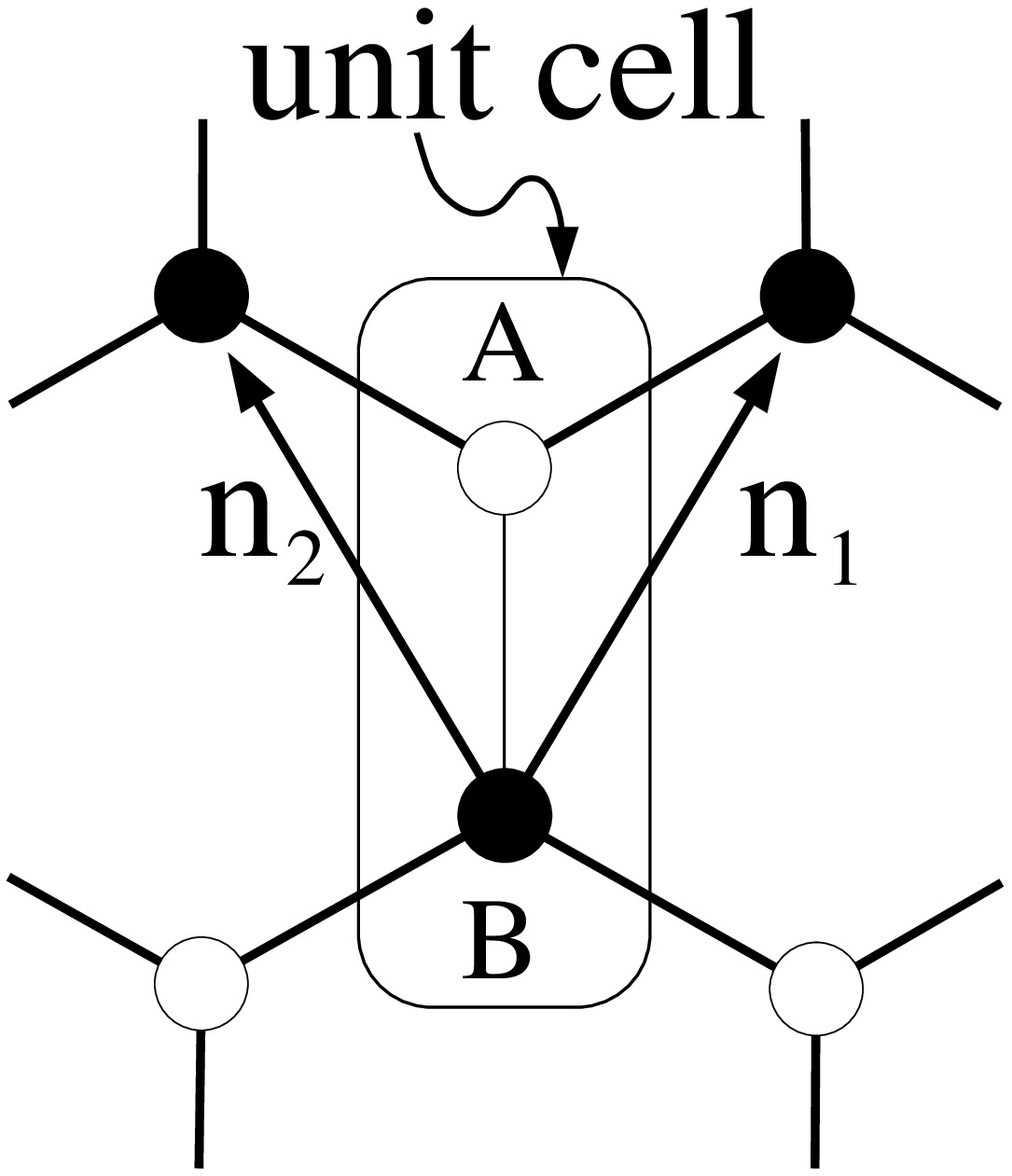,width=0.8\linewidth}
\end{minipage}
\begin{minipage}[b]{0.5\linewidth}
(a)
\end{minipage}
\begin{minipage}[b]{0.35\linewidth}
(b)
\end{minipage}
\caption{\label{unitcell}(a) Labelling ($x,y,z$) of the three bond types in a honeycomb lattice, and site numbering $1$ to $6$ for plaquette operator $W_{\hexagon}$. (b) The unit cell, lattice vectors and sublattice convention used throughout this work.}
\end{figure}
\section{The Kitaev honeycomb model}
\label{TheModel}

The Kitaev honeycomb model 
consists of 
spins on the sites of a honeycomb lattice, with nearest neighbour interactions that 
depend on the orientation of the bond, labelled $x$, $y$ or $z$ as shown in Fig.~\ref{unitcell}(a).
Representing a spin variable at site $j$ by the Pauli matrices $\tilde{\sigma}_{j}^{\alpha}$ and denoting the neighbouring sites by $k$, the Hamiltonian reads
\begin{equation}
\widetilde{H} = -\dsum_{\text{x-links}}J_{x}\tilde{\sigma}_{j}^{x}\tilde{\sigma}_{k}^{x}- \dsum_{\text{y-links}}J_{y}\tilde{\sigma}_{j}^{y}\tilde{\sigma}_{k}^{y}- \dsum_{\text{z-links}}J_{z}\tilde{\sigma}_{j}^{z}\tilde{\sigma}_{k}^{z}.
\end{equation}
In the following we indicate the bond orientation between sites $j$ and $k$ using $\alpha_{jk} = x, y$ or $z$. Without loss of generality, we take $J_\alpha \geq 0$.

\subsection{Mapping to free fermions}
\label{MappFermions}

This
Hamiltonian is exactly solvable, for example with a local transformation $\widetilde{\sigma}^{\alpha}_{j} = ib_{j}
^{\alpha}c_{j}$ which represents each spin using the four Majorana fermions \cite{Kitaev:2006ys} $b_{j}^x,b_{j}^y,b_{j}^z$ and $c_{j}$. This transformation enlarges the Hilbert space and to emphasize this difference we mark operators in the Hilbert space of spins with a tilde and those in the space of Majoranas without. After this transformation the Hamiltonian reads
\begin{equation}
H_{\hat{u}} = \dfrac{i}{2}\dsum_{j,k}J_{\alpha_{jk}}\hat{u}_{jk}c_{j}c_{k}\qquad \hat{u}_{jk} = ib_{j}^{\alpha_{jk}}b_{k}^{\alpha_{jk}}.
\label{quadH}
\end{equation}
The operators $\hat{u}_{jk}$ commute with each other and with $H_{\hat{u}}$.  One can therefore fix the values of $\langle\hat{u}_{jk}\rangle =u_{jk}= \pm1$, move to a subspace of the full Hamiltonian and obtain a bilinear form in the $c_j$'s. 
For each set $\{u_{jk}\}$, the resulting Hamiltonian $H_u$ inherits a bipartite structure from the honeycomb lattice. This is displayed by introducing, for a lattice of $N$ unit cells as depicted in Fig.~\ref{unitcell}(b), two $N$-component vectors of Majorana fermion operators, $c_{A}$ and $c_{B}$, from sublattices $A$ and $B$ respectively, and an $N\times N$ matrix $M$, with entries $J_{\alpha_{jk}}{u}_{jk}$. Then 
\begin{equation}
H_{u}=\frac{i}{2}\left(\begin{array}{cc}c^{T}_{A}&c^{T}_{B}\end{array}\right)\left(\begin{array}{cc} 0 & M \\ -M^{T} & 0\end{array}\right)\left(\begin{array}{c}c_{A}\\c_{B}\end{array}\right)\,.
\label{hopC}
\end{equation}
The energy levels of $H_u$ can be expressed in terms of the eigenvalues $\pm S_m$ of the $2N\times 2N$ matrix
\begin{equation}
H = \left(\begin{array}{cc} 0 & M \\ M^{T} & 0\end{array}\right)\,.
\label{hopA}
\end{equation}
The matrix $H$ can be interpreted as a tight binding model and we will use this
viewpoint extensively. The connection between the eigenvalues of $H$ and those of $H_u$ is
as follows.
The matrix $M$ has a singular value decomposition $M = USV^{T}$, where $U$ and $V$ are $N\times N$ orthogonal matrices and $S$ is an $N \times N$ positive semidefinite diagonal matrix. 
Label a unit cell by $\rr$ and let the two Majorana fermion operators within it be $c_{A,\rr}$ and $c_{B,\rr}$. 
Then Majorana fermions in the eigenbasis are
\begin{eqnarray}\label{eigenmajorana}
c_{m,A} &=& \sum_{\rr} U^{T}_{m\rr}c_{A,\rr} \nonumber\\
{\rm and} \qquad c_{m,B} &=& \sum_{\rr} V^{T}_{m\rr}c_{B,\rr} \,.\qquad
\end{eqnarray}
Defining the complex fermions $a_{m} = \frac{1}{2}\left(c_{m,A}+ic_{m,B}\right)$ the Hamiltonian 
is brought into the diagonal form
\begin{equation}
H_{u} = {\rm i}\dsum_{m = 1}^{N}S_{m}c_{m,A}c_{m,B} \equiv
\dsum_{m = 1}^{N}S_{m}\left(2a_{m}^{\dagger}a_{m}-1\right).
\label{HuDiag}
\end{equation}

The three Hamiltonians presented so far, $\widetilde{H}, H_{\hat{u}}$ and $H_{u}$, 
act in different Hilbert spaces and should not be confused. For instance, their respective Hilbert space dimensions are $2^{2N}$, $\sqrt{2}^{4\times 2N} = 2^{4N}$ and $2^{N}$. 

\subsection{Emergent gauge field and non-dynamical fluxes}
\label{EmergField}

Kitaev showed that there exist non-dynamical, commuting flux operators $\widetilde{W} = \tilde{\sigma}_{j}^{\alpha_{jk}}\tilde{\sigma}_{k}^{\alpha_{jk}}\tilde{\sigma}_{k}^{\alpha_{kl}}\tilde{\sigma}_{l}^{\alpha_{kl}}\tilde{\sigma}_{l}^{\alpha_{lm}}\tilde{\sigma}_{m}^{\alpha_{lm}}\cdots\tilde{\sigma}_{n}^{\alpha_{nj}}\tilde{\sigma}_{j}^{\alpha_{nj}}$ defined along any closed loop on the lattice. They have eigenvalues $\pm1$, and so are $Z_{2}$ variables. Moreover, their
effect on fermion hopping is that of a flux through the corresponding plaquette. With periodic boundary conditions there are two independent non-contractible loops winding around the system. A complete and independent set of variables specifying the flux state is provided by the values of all but one of the fluxes through the hexagonal plaquettes, supplemented by those of the pair of fluxes through the non-contractible loops. 

Indeed, these fluxes are simply related to the variables $u_{jk}$ encountered above. Numbering the sites around a plaquette from 1 to 6 [see Fig. \ref{unitcell}(a)] the Z$_2$ flux through a plaquette can be written as $W_{{\epsfig{file=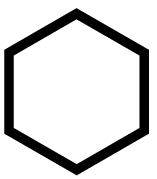,width=0.2cm}}} = u_{21}u_{23}u_{43}u_{45}u_{65}u_{61}$.  The spectrum of $H_u$ depends only on these fluxes \cite{Kitaev:2006ys} but because of invariance under gauge transformations, many choices of the set $\lbrace u_{jk} \rbrace$ encode the same flux sector. The simplicity of the model is that the gauge field represented by $u_{jk}$ has no dynamics.

The appearance of a gauge degree of freedom is a consequence of the transformation to Majorana operators, which doubles the dimension of the Hilbert space for each spin. Using a variable $\theta_{j} = \pm1$ on each lattice site, the gauge transformations are implemented in the standard way, as $c_{j}\rightarrow\theta_{j}c_{j}$ and $u_{jk}\rightarrow \theta_{j}u_{jk}\theta_{k}$. Due to the doubling in Hilbert space dimensions per spin, a projection operator must be applied to eigenstates of $H_{\hat{u}}$ in order to obtain those of $\tilde{H}$.
In fact, however, for the operators considered in this paper, matrix elements calculated using eigenstates of $H_{\hat{u}}$ are the same as those calculated using the projected physical states. A full discussion of the projection operation is given in Appendix \ref{ProjectionAppend}.

\subsection{Excitations: fluxes and fermions}
\label{Excitations}

The ground state energy in a particular flux sector is, from Eq.~(\ref{HuDiag}),  $E_{0} = -\sum_{m}S_{m}$,
and the absolute ground state lies in the flux sector that minimises this energy.
There are two distinct types of excitation. First, within the same flux sector, ``matter'' excitations involve the occupancy of the fermionic modes, $(2a^\dagger_m a_m -1) = 1$ in Eq.~(\ref{HuDiag}). Second, flux excitations consist of a set of flux values different from those in the ground state.
For the disorder-free system, the ground state is flux free.\cite{Lieb:1994yq} As a function of the interaction parameters $J_\alpha$, there are four phases in the absence of an applied field: three symmetry-equivalent phases in which matter excitations are gapped, when one exchange dominates (e.g. $J_{z} > J_{x}+J_{y}$), and one phase in which they are gapless, around the ``isotropic'' point  $J_{x}{=}J_{y}{=}J_{z}{=}J$.
Flux excitations are generically gapped: their cost is given by the difference in fermionic ``zero-point'' energies, $E_0$, between excited and ground state flux sectors.

\section{Introducing vacancies}
\label{ModelingVacancies}
 
Vacancies have a dramatic effect on local properties of the Kitaev model. We will show that they bind a flux in the ground state. Furthermore, the removal of a site reduces the number of ``constraints'' on the spins adjacent to it and allows a magnetic moment to form. The formation of this moment, its susceptibility and the interaction between vacancy moments are the subject of this paper. 
 
Our analysis starts from the observation that the steps described above (representing spins using Majorana fermions and fixing the values of $u_{jk}$) remain valid in the presence of both vacancies and the leading terms in the Zeeman energy. The effect of a vacancy manifests itself on several levels in the description of the model. 

\begin{figure}[ht]
\epsfig{file=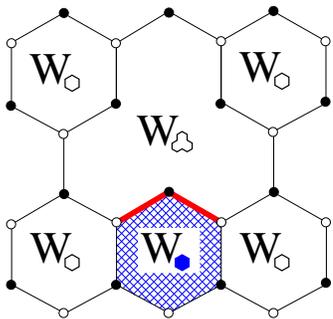,width = 0.5\linewidth}
\caption{In the gapped phase (vertical bonds $J_z > J_x + J_y$) the ground state sector has a $Z_2$
flux through the vacancy plaquette.}
\label{Fluxes}
\end{figure}

First, the three original plaquettes that meet at the vacancy site now form one big plaquette (different from all other plaquettes: see Fig.~\ref{Fluxes}) and hence the number of independent fluxes decreases by two. Second, and as a consequence, there is a qualitative change in the way that a Zeeman field couples to spin system. A key result we reach is a derivation of a modified version of  the tight binding model, Eq. (\ref{hopA}), valid in the presence of vacancies.

To set the context it is useful to recall the effect of a Zeeman field in the model without vacancies.\cite{Kitaev:2006ys}
On including the field, the Hamiltonian no longer commutes with the Z$_{2}$ fluxes $\widetilde{W}$, which therefore become dynamical. This spoils solvability but a weak field may be treated perturbatively. Since there is a finite energy gap $\Delta_{W}$ between the ground state and other flux sectors, at field strength $h\ll \Delta_{W}$ one can project onto that flux sector. Practically all our results presented here thus pertain to a particular (typically the ground-state) flux sector.

In more detail, for the undiluted lattice, matrix elements of the Zeeman energy between states from the same flux sector are all zero (see Ref.~\onlinecite{Kitaev:2006ys} and Appendix~\ref{ProjectionAppend}). The leading contribution to a projected Zeeman Hamiltonian is therefore second order in $h$, which results in a non-vanishing but finite magnetic susceptibility at small $h$.\cite{Kitaev:2006ys}

\begin{figure}
\begin{tabular}{c@{\qquad}c}
\epsfig{file=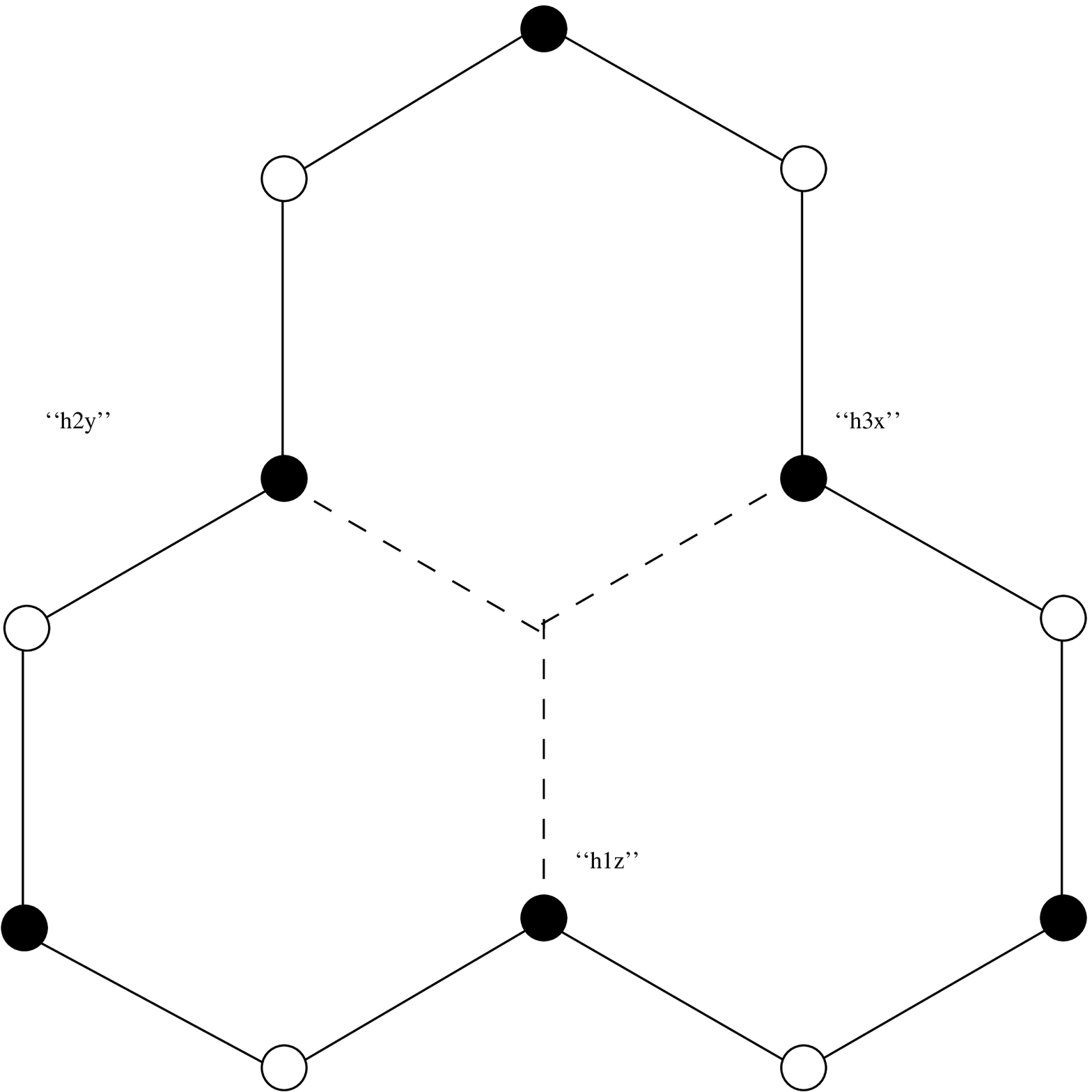, width=3.5cm}&\epsfig{file=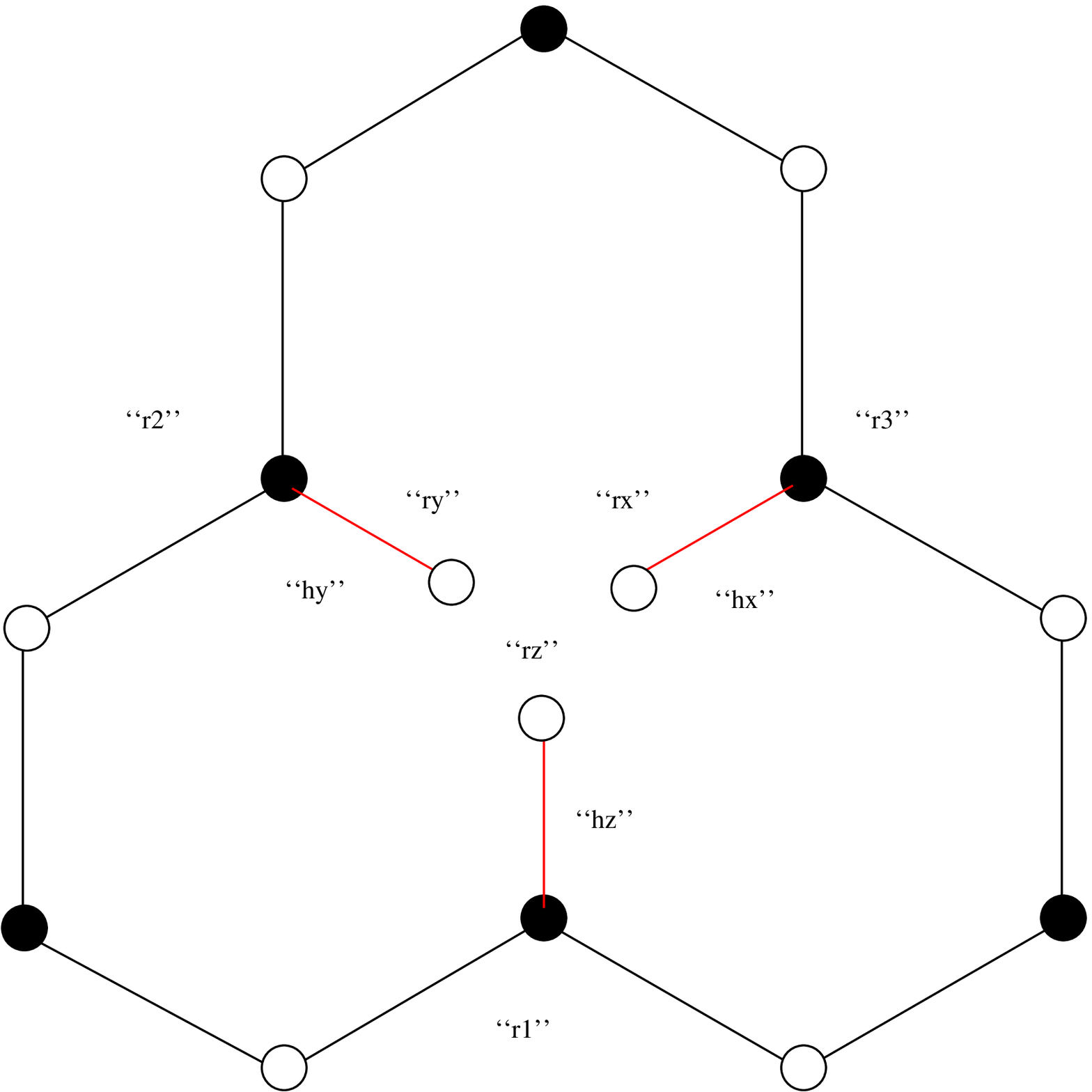, width=3.5cm}\\
(a)&(b)
\end{tabular}
\caption{\label{EffHam}(a) Sites at which field components $h_x$, $h_y$, and $h_z$ contribute linearly to the projected Hamiltonian with a vacancy.  (b) Representation as an equivalent tight binding model, with three new sites coupled with respective hopping matrix elements $h_x,h_y,h_z$.}
\end{figure}

The altered nature of the Zeeman coupling comes about because, with the merging of three plaquettes into one, four
previously distinct flux sectors collapse into a single one. As a result, some terms in the Zeeman Hamiltonian now commute with all fluxes $\widetilde{W}$ in the model with vacancies:
the non-zero matrix elements of these terms connect eigenstates of $\widetilde{H}$ that belong to different flux sectors in the undiluted system but to the same sector in the diluted system. They thus  generate contributions to the projected Zeeman Hamiltonian that are linear in $h$, and these are responsible for the dominant local magnetic response at weak field.
Using the labelling indicated in Fig.~\ref{EffHam}(a), this part of the Zeeman energy is
\begin{equation}\label{ProjZeemanEn}
\widetilde{H}_{\rm Z} = -(h_x \tilde{\sigma}^x_1 + h_y \tilde{\sigma}^y_2 + h_z \tilde{\sigma}^z_3)~.
\end{equation}

To show that the Hamiltonian $\widetilde{H}+\widetilde{H}_{Z}$ can be diagonalised using the steps set out in Sec. \ref{TheModel}, we find the counterpart to Eq.~(\ref{hopC}) arising from $\widetilde{H}_{Z}$. To start, recall that each bond of the lattice is associated with two Majorana fermions. As removal of a site breaks three bonds, it leaves three unpaired Majorana fermions. We denote them by $b^\alpha_j$, where $j=1$, 2 or 3 labels the sites adjacent to the vacancy.
With this notation, the contribution to $H_{\hat{u}}$ generated by  the term $h_{\alpha}\tilde{\sigma}^{\alpha}_{j}$ in $\widetilde{H}_{Z}$ is $ih_{\alpha} b^{\alpha}_{j}c_{j}$. The Majorana fermions $b^\alpha_j$ are represented 
in a tight binding model by three orbitals that do not appear for the undiluted system. These orbitals are coupled to the ones representing the $c_j$'s by hopping of strength $h_{\alpha}$, as shown in Fig.~\ref{EffHam}(b).  

In studying this and related problems, it will be useful to keep in mind some basic properties
of tight binding models with nearest hopping on bipartite lattices: with $N_A$ ($N_B$) orbitals on the A (B) sublattice, energy eigenvalues generically consist of $|N_A-N_B|$ zero modes and $\min\{N_A,N_B\}$ pairs (the two members of a pair having energies of equal magnitude and opposite sign). 

The task then is to calculate the field dependence of the ground state energy of the Kitaev model with one or more vacancies, using the tight binding model of Fig.~\ref{EffHam}(b), and its generalisation in the case of more than one vacancy.
We do this by using a T-matrix approach [see Eq.~(\ref{t-matrix})] to express 
properties of the system with vacancies in terms of the Green function of the 
hexagonal lattice tight binding model, for which there are convenient analytic expressions in
the flux-free sector (reviewed in Appendix B). 
We will find that there are marked differences between the gapped and gapless phases and 
we separate our analysis accordingly. 


\section{Gapped Phase}
\label{GappedPhase}

We begin our discussion with the gapped phase, whose relatively simple structure allows us to make considerable progress starting from the single vacancy problem. Indeed its very name holds the promise of a perturbative treatment, and for $J_z > J_x + J_y$, we can use $j_{x,y} = J_{x,y}/J_z$ as small parameters in an expansion.

Our analysis proceeds in several steps. We first derive an effective Hamiltonian demonstrating that vacancies in the gapped phase bind a flux. Next, we provide a detailed analysis of the single vacancy problem, where a lone fermionic zero mode appears in the energy gap. This zero mode is localised in real space in a striking fashion: its probability density is zero outside a wedge that has the vacancy at its apex, and decays exponentially with distance from the apex. This mode we show carries an effective paramagnetic moment on the site linked to the vacancy by a strong bond; the size of the moment grows with decreasing $j_{x,y}$.

Zero modes belonging to vacancies on different sublattices ``hybridise'' when their wedges overlap, as we demonstrate by an analysis of the two-vacancy problem. For this, we can derive an effective Hamiltonian describing the energetics in the gap in the presence of a field.

Finally, the many vacancy problem leads us to consider an effective bipartite random hopping problem (BRH). This can be analysed in the spirit of a strong disorder renormalisation group treatment, as nearby pairs of vacancies hybridise exponentially more strongly than distant ones. This leads us to a  strongly divergent low energy density of states near $E = 0$. 

The remainder of this section gives a detailed account of this set of phenomena.

\subsection{Flux Binding}\label{flux binding}
The ground state flux sector of the undiluted Kitaev model is flux free, for example with $u_{jk} = +1$ on every link. However, as we now show, the removal of a site binds a flux to the vacancy plaquette. 

The energy differences between flux sectors may be found at small $j_{x,y}$ without resorting to the Majorana decomposition given above.\cite{Kitaev:2006ys}  Instead we extend to a system with vacancies the approach originally presented by Kitaev for the undiluted model.  To this end, write the Hamiltonian as $\widetilde{H} = \widetilde{H}_{0}+\widetilde{V}$, where 
\begin{equation}\label{H0andZeeman}
\begin{array}{rl}
\widetilde{H}_{0} &= -J_{z}\dsum_{\text{z-links}}\tilde{\sigma}_{j}^{z}\tilde{\sigma}_{k}^{z}\\\\
\widetilde{V} &= -J_{x}\dsum_{\text{x-links}}\tilde{\sigma}_{j}^{x}\tilde{\sigma}_{k}^{x}-J_{y}\dsum_{\text{y-links}}\tilde{\sigma}_{j}^{y}\tilde{\sigma}_{k}^{y}\,.
\end{array}
\end{equation}
The ground state of $\widetilde{H}_{0}$ is $2^{N}$-fold degenerate and has an energy $E_{0} = -J_{z}N$.  The low energy states of  $\widetilde{H}$ can be understood by projecting onto ground states of $\widetilde{H}_{0}$ and working  perturbatively in $\widetilde{V}$. In this subspace $\widetilde{V}$ takes the form of an effective flux Hamiltonian that acts within the ground states of $\widetilde{H}_{0}$, lifting the degeneracy.  

We use standard perturbation theory to find the effective flux Hamiltonian,\cite{Bloch:1958} denoting the $n^{\text{th}}$ order perturbation by $\widetilde{H}^{(n)}_{\text{eff}}$ and with $\Pi$ a projection onto ground states of $\widetilde{H}_{0}$. Further, let $|a\rangle$ and $E_a$ be the eigenstates and energy levels of $\widetilde{H}_{0}$. The action of $\widetilde{V}$ on such an eigenstate is to flip two spins and thus change the energy by $4J_{z}$. The first two terms in the perturbation theory are
\begin{equation}
\begin{array}{rl}
\widetilde{H}^{(1)}_{\text{eff}} =& \Pi \widetilde{V} \Pi = 0\\[2ex]
\widetilde{H}^{(2)}_{\text{eff}} =& \Pi \dsum_{a} '\dfrac{\widetilde{V}|a\rangle\langle a|\widetilde{V}}{E_{0}-E_{a}} \Pi =  -\dfrac{NJ_{z}}{4}\left(j_{x}^{2}+j_{y}^{2}\right)
\end{array}
\end{equation}
where the primed summation is over states outside the ground state manifold of $\widetilde{H}_{0}$. 
At general order, $\widetilde{H}^{(2n-1)}_{\text{eff}}$ is zero 
and the most important contribution to $\widetilde{H}^{(2n)}_{\text{eff}}$ is
\begin{equation}
\Pi \dsum_{a,b,\dots,2n-1} '\dfrac{\widetilde{V}|a\rangle\langle a|\widetilde{V}|b\rangle\langle b|\widetilde{V}\cdots \widetilde{V}|2n-1\rangle\langle 2n-1|\widetilde{V}}{(E_{0}-E_{a})(E_{0}-E_{b})\cdots(E_{0}-E_{2n-1})}\Pi \,.
\end{equation}
Here we omit for conciseness other terms in $\widetilde{H}^{(2n)}_{\text{eff}}$ that have a subleading effect on degeneracies. Without dilution, the lowest order term that reduces the ground state degeneracy is 
\begin{equation}
\widetilde{H}^{(4)}_{\text{eff}} = -\dfrac{J_{z}}{16}j_{x}^{2}j_{y}^{2}\dsum \widetilde{W}_{\epsfig{file=FluxHex.eps,width=0.2cm}}+\text{const}\,,
\end{equation}
where $\widetilde{W}_{\epsfig{file=FluxHex.eps,width=0.2cm}}$ is the flux through a hexagonal plaquette of the lattice. The prefactor $-\frac{1}{16}$ is found from summing $4!$ terms, corresponding to the permutations of elements from the four $\widetilde{V}$ operators. 

With dilution,
at each vacancy there is one hexagonal plaquette that has a larger prefactor, indicated by $\epsfig{file=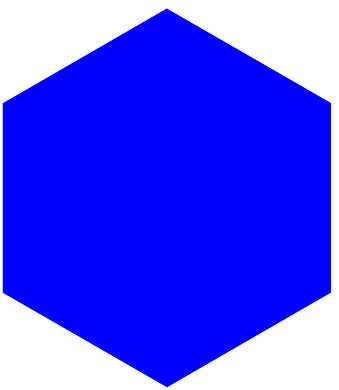,width=0.25cm}$ in Fig. \ref{Fluxes}. Here, in place of a spin pair with strong coupling $J_z$, there is simply a lone spin. In consequence, when the two elements of $\widetilde{V}$ adjacent to the vacancy plaquette act, the energy change is only $2J_{z}$. These elements are shown in Fig. \ref{Fluxes}, where we also define the vacancy plaquette $\epsfig{file=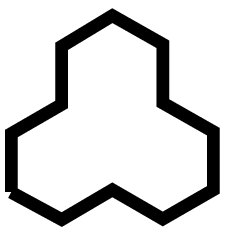,width=0.3cm}$.

Perturbation theory must be extended to $8^{\text{th}}$ order to find a term that depends on the flux through the vacancy plaquettes. This involves a summation of $8!$ permutations from terms encircling $\epsfig{file=FluxVac.eps,width=0.3cm}$ and an additional $8!$ from those encircling both $\epsfig{file=FluxVac.eps,width=0.3cm}$ and $\epsfig{file=FluxHexVac.eps,width=0.25cm}$ plaquettes. 
The resulting Hamiltonian for an isolated vacancy reads
\begin{align}
\widetilde{H} \simeq&~ E_{0}-J_{z}~j_{x}^{2}j_{y}^{2}\left(\dfrac{1}{16}\dsum \widetilde{W}_{\epsfig{file=FluxHex.eps,width=0.2cm}}+\dfrac{3}{8}\dsum \widetilde{W}_{\epsfig{file=FluxHexVac.eps,width=0.2cm}}\right)\nonumber\\
&+J_{z}~j_{x}^{4}j_{y}^{4}\left(\dfrac{21}{2^{10}}\dsum \widetilde{W}_{\epsfig{file=FluxVac.eps,width=0.2cm}}-\frac{33}{2^{11}}\dsum \left(\widetilde{W}_{\epsfig{file=FluxVac.eps,width=0.2cm}}\times\widetilde{W}_{\epsfig{file=FluxHexVac.eps,width=0.2cm}}\right)\right)\,.
\label{EffFluxHam}
\end{align}
The fluxes $\widetilde{W}_{\epsfig{file=FluxHex.eps,width=0.2cm}}$ and $\widetilde{W}_{\epsfig{file=FluxHexVac.eps,width=0.2cm}}$ are determined by the fourth order terms in Eq. (\ref{EffFluxHam}). Both the associated plaquettes are flux free in the ground state: $\langle\widetilde{W}_{\epsfig{file=FluxHex.eps,width=0.2cm}}\rangle = \langle\widetilde{W}_{\epsfig{file=FluxHexVac.eps,width=0.2cm}}\rangle = +1$. In such states the terms involving $\widetilde{W}_{\epsfig{file=FluxVac.eps,width=0.2cm}}$ can be combined to give an energy contribution $(9/2^{11})\,J_{z}j_{x}^{4}j_{y}^{4}\sum\widetilde{W}_{\epsfig{file=FluxVac.eps,width=0.2cm}}$. The positive coefficient of $\sum \widetilde{W}_{\epsfig{file=FluxVac.eps,width=0.2cm}}$ indicates that the introduction of vacancies changes the ground state flux sector: each vacancy binds a flux, so that in the ground state $\langle \widetilde{W}_{\epsfig{file=FluxVac.eps,width=0.2cm}}\rangle = -1$. 
This approach may be extended to include larger voids: the flux through those larger voids makes a contribution to the effective flux Hamiltonian that is higher order in $j_{x}$ and $j_{y}$. Note that the resulting flux gaps are numerically rather small. For $j_x = j_y = 1/3$, one obtains a number of order $10^{-6} J_z$ for a single vacancy plaquette. The theory developed in the remainder of this section needs to be understood as applying for fields small on this scale.

\subsection{A single vacancy}
Having seen that a vacancy binds a flux, we now discuss its magnetic properties. 
For this purpose it is convenient to switch techniques, from perturbation theory for the spin Hamiltonian to the
Majorana fermion representation introduced in Section \ref{MappFermions}.
We show that the vacancy generates a fermionic zero mode, which in turn leads to a twofold degeneracy of all eigenstates of $\widetilde{H}$.
The degeneracy is lifted by  non-zero Zeeman field $\hh$, 
reflecting the formation of a free moment distributed over sites close to the vacancy. The moment has an anisotropic effective $g$-factor that depends on the parameters $j_{x,y}$.
 
We first recall the gapped spectrum of matter excitations for the undiluted lattice, which can be found by Fourier transform of the tight-binding Hamiltonian, Eq.~(\ref{hopA}). We define basis vectors $\nn_1$ and $\nn_2$ for the honeycomb lattice as in
Fig.~\ref{unitcell}(a), and introduce a wavevector $\qq$ with components $q_{1} = \qq.\nn_{1}$ and $q_{2} = \qq.\nn_{2}$.  The eigenstates of  $H$ with this wavevector have energies $\pm S_{\qq} = \pm|J_{z}+J_{x}e^{iq_{1}}+J_{y}e^{iq_{2}}|$: these form two bands, arranged symmetrically around zero. The minimum of $S_{\qq}$ over $\qq$ in the phase under discussion ($0\leq j_x,j_y < 1$) is $\Delta = J_{z}-J_{x}-J_{y}$, yielding a band gap for $H$ of $2\Delta$.

\begin{figure}
\begin{center}
\begin{tabular}{cc}
\epsfig{file=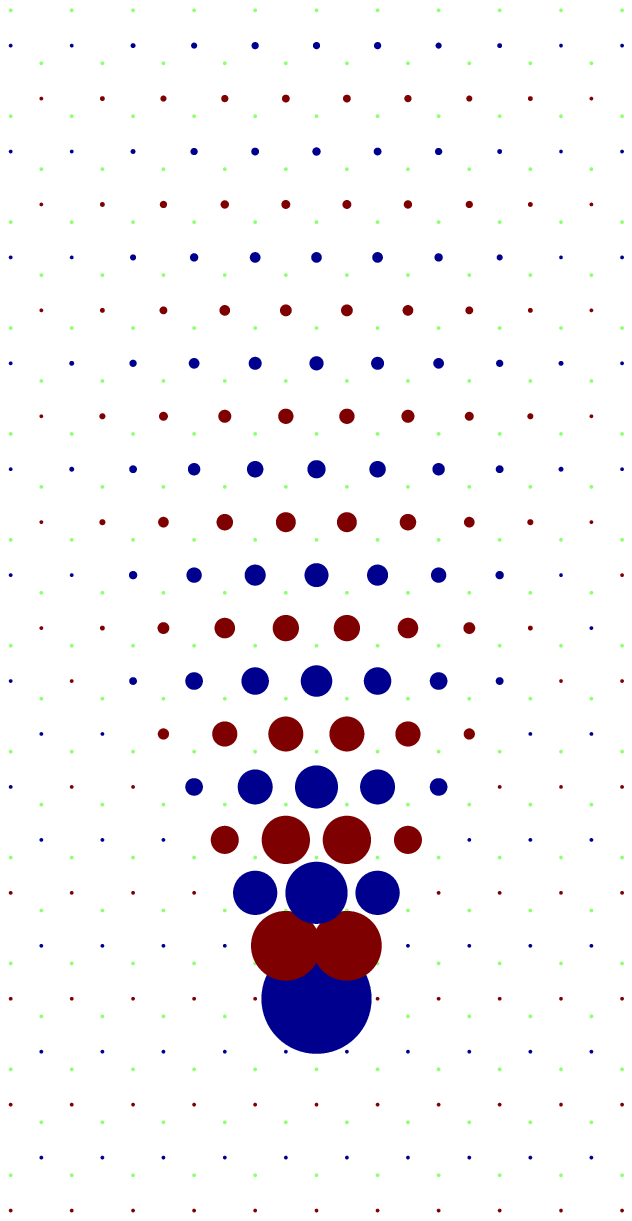,width=0.2\textwidth}&\epsfig{file=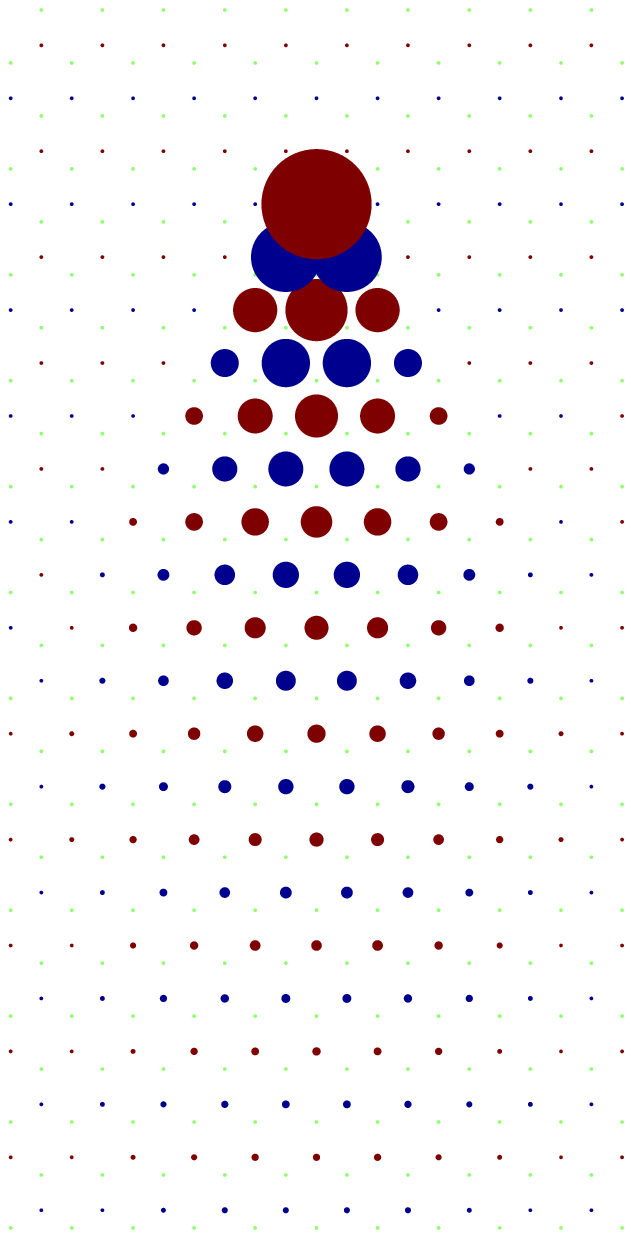,width=0.2\textwidth}\\
\text{A vacancy}&\text{B vacancy}
\end{tabular}
\end{center}
\caption{\label{1vacZeroMode}Vacancy zero modes from the tight binding model, Eq.~(\ref{hopA}). Area and colour of each circle respectively represent magnitude and sign of the wavefunction at that site.}
\end{figure}

A single vacancy on one sublattice generates a zero mode,  an eigenstate of the honeycomb lattice tight binding model at zero energy, with amplitude only on the opposite sublattice. In the gapped phase the eigenfunction is exponentially localised and has a particularly simple form: sites with non-zero amplitude are located inside a $60^{\circ}$ wedge emerging from the vacancy and in the direction of the strong bonds. 
The zero mode wavefunction has a straightforward expression in a system without flux. In this case, for a B vacancy at the origin, its amplitude on the A site in unit cell $\rr$ is a representation of Pascal's triangle, with 
\begin{equation}
\begin{array}{c} \Psi^{B}(\rr) = \n (-1)^{n_{1}+n_{2}}~j_{x}^{n_{1}}~j_{y}^{n_{2}}{n_{1}+n_{2} \choose n_{1}}
\label{BVacZeroMode}
\end{array}
\end{equation}
if the site lies within the wedge ($n_{1}, n_{2} \geq 0$). Thus, inside the wedge the amplitude on A sites decays with distance from the vacancy in a direction-dependent fashion. Outside this wedge, and on all B sites, the amplitude is zero. The wavefunction associated with a vacancy on the opposite sublattice is related to this one by inversion symmetry. Both cases are illustrated in Fig.~\ref{1vacZeroMode}.
The normalisation constant 
\begin{equation}
\n  = \sqrt[4]{\left(1-j_{x}^{2}-j_{y}^{2}\right)^{2}-4j_{x}^{2}j_{y}^{2}}
\label{NormConst}
\end{equation}
will play in important role in what follows. In a system with a single flux bound to an isolated vacancy, the zero mode wavefunction has site amplitudes of the same magnitude as just described. The concomitant ($Z_2$) phases are of course gauge-dependent. 

The presence of a zero mode is linked to the formation of a free local moment around the vacancy, and this moment is polarised by the local field components that are included in projected Zeeman energy, Eq.~(\ref{ProjZeemanEn}). 
To examine moment formation in detail, we compute the field-dependence of the ground state energy of the Kitaev model with a vacancy, using the tight binding model of Fig.~\ref{EffHam}(b). As discussed, the basis orbitals for this tight binding Hamiltonian consist of all those appearing in the honeycomb lattice with a vacancy, together with three additional ones arising from unpaired Majorana fermions $b_j^\alpha$. 

At $h = 0$, the matrix elements of the tight binding Hamiltonian involving these additional orbitals vanish. Its spectrum in this case therefore includes four zero modes 
located in the middle of the gap between positive and negative energy bands. 
The zero-mode subspace is spanned by the orbitals $r_x$, $r_y$, $r_z$ and the wavefunction $\Psi(r)$ (see Fig.~\ref{EffHam}(b) and Eq. (\ref{BVacZeroMode})).

At leading order $h$ acts within this subspace, lifting the degeneracy of the zero modes. In fact, since the vacancy mode $\Psi(\rr)$ has no amplitude on sites $\rr_{1}$ and $\rr_{2}$, the orbitals $r_x$ and $r_y$ are unaffected, ultimately yielding states with energies quadratic in $h_x$ and $h_y$ via coupling to the finite energy bands. 

By contrast, states arising from $\Psi(\rr)$ and $r_z$ have energies linear in $h_z$ for small $h$.
Since these states lie within the energy gap, it is natural to project onto the two-dimensional subspace
that they span.
The projected tight binding Hamiltonian has the form
\begin{equation}\label{Heff1vac}
\left(\begin{array}{cc}0&\n  h_{z} \\ \n  h_{z} &0\end{array}\right),
\end{equation}
where $\n$ appears as the amplitude $\Psi(r_3)$ on the site adjacent to the vacancy.
Viewing the projection in terms of Majorana fermions (taking for definiteness the case
of a B vacancy at the origin) we have [following Eq.~(\ref{eigenmajorana})] 
\begin{equation}
c_{1,A} = \sum_{\rr} \Psi^B(\rr) c_{A,\rr} \quad {\rm and} \quad c_{1,B} = b^z_{\rr_3}\,.
\end{equation}
The Hamiltonian $H_{u}$ [Eq.~(\ref{HuDiag})] projected onto the low energy states can then be written as 
\begin{equation}\label{vac-fermion}
H_u = S_1(2a^\dagger_1a_1 -1) 
\,,
\end{equation}
with $S_1 = \n |h_{z}|$. 
The ground state magnetisation $m_{z} = -\partial_{h_{z}}E_{0}$ is thus 
$\n {\rm sgn}(h_z)$ at leading order in $h$. Moreover, $H_{u}$ has a low-lying excited state,
with the opposite magnetisation, which is higher in energy by $2 \n |h_z|$, formed by 
taking the occupation number $a^\dagger_1 a_1$ to be one rather than zero.
 
We conclude that the vacancy has generated a paramagnetic moment of size $\n$. 
A consequence of the form of the projected Zeeman energy $\widetilde{H}_{\rm Z}$ is that the magnetisation associated with the moment is entirely localized on the site $\rr_{3}$ adjacent to the vacancy. 
Indeed, for zero $j_{x},j_{y}$, the Kitaev model with a vacancy consists a number of spin pairs strongly coupled by $J_z$ exchange, together with one unpaired spin on this site. However, even in the limit of $j_x, j_y$ small, the local moment is strikingly different from a free spin because the limits of small $j$ and small $h$ do not commute.
In particular, irrespective of field orientation,  only the $z$-component of the moment develops 
a finite expectation value at small $\hh$, because all matrix elements of $\tilde{\sigma}^x_{\rr_{3}}$ and $\tilde{\sigma}^y_{\rr_{3}}$ within the zero-mode subspace vanish.  
Equivalently, the local moment has an anisotropic $g$-tensor, with
$g_{zz}$ as the only non-zero component, varying with $j_x$ and $j_y$ between $g_{zz}=0$ 
on the phase boundary with the gapless phase, to $g_{zz} = 1$ for $j_{x},j_{y}\rightarrow 0$.

This non-trivial form of the $g$-tensor reflects the fact that the local moment describes a collective coordinate. Some further insight comes from considering a system in which, additionally, the spin at $\rr_{3}$ has been removed, leaving vacancies on two sites adjacent in the $z$-direction. Without Zeeman terms, eigenstates of this spin model can be shown (we omit details) to have a two-fold degeneracy that arises from the double vacancy, and we use $|+\rangle$ and $|-\rangle$ to denote two chosen othonormal ground state wavefunctions.

We employ the states $|\pm \rangle$ to construct the two lowest-energy eigenfunctions, $| 0 \rangle$ and $| 1 \rangle$, of single vacancy model including a weak projected Zeeman field.
Introducing the eigenstates $|\uparrow\rangle$ and $|\downarrow\rangle$ of $\sigma^z_{\rr_{3}}$, these 
may be written as
\begin{eqnarray}
|0\rangle &=& \sqrt{(1+{g_{zz}}/{2}})|\uparrow\rangle \otimes |+\rangle + \sqrt{(1-g_{zz}/2)}|\downarrow\rangle \otimes |-\rangle\nonumber\\
{\rm and}&&\nonumber\\
|1\rangle &=& \sqrt{(1-g_{zz}/2)}|\uparrow\rangle \otimes |-\rangle + \sqrt{(1+{g_{zz}}/{2}})|\downarrow\rangle \otimes |+\rangle\nonumber
\end{eqnarray} 
Then $\langle 0| \sigma^z_{\rr_{3}}|0\rangle = g_{zz}$ and $\langle 1| \sigma^z_{\rr_{3}}|1\rangle = -g_{zz}$, but for any state in this subspace $\langle \sigma^x_{\rr_3}\rangle=\langle \sigma^y_{\rr_3}\rangle=0$.



\subsection{Two vacancies}
We next consider a pair of vacancies. Our main interest is in the interaction between the local moments formed near well-separated vacancies, and in particular we assume a sufficiently large separation that they are associated with different hexagons of the lattice. Then they generate two separate vacancy plaquettes of the type shown in Fig.~\ref{EffHam}. In the Majorana fermion representation of the Kitaev Hamiltonian, Eq.~(\ref{quadH}), two vacancies lead to six uncoupled Majorana fermions from the six broken bonds. In the tight binding model, Eq.~(\ref{hopA}), these are represented as six uncoupled zero energy orbitals. At large vacancy separation, however, the leading Zeeman coupling involves only the $z$-components of field for $j_x,j_y <1$, and so only two of these orbitals play an important role \cite{footnote:I}. There are also two localised modes similar to that of Eq.~(\ref{BVacZeroMode}), and so we must consider a total of four states in the low-energy subspace.

The resulting behaviour depends crucially on whether the vacancies belong to the same or opposite sublattices. For vacancies on the same sublattice, the localised modes are both at zero energy. In this case, therefore, two vacancies produce two essentially independent paramagnetic moments, each similar to that for a single vacancy.

By contrast, vacancies on opposite sublattices may give rise to moments that interact. More specifically, consider two such vacancies, placed so that each one lies inside the zero-mode wedge of the other. In this case the 
vacancy modes of the tight binding model hybridise, forming a pair of eigenstates at energies $\pm \varepsilon$, with $\varepsilon \ll \Delta$ if
the vacancy separation is large. As in our discussion of a single vacancy, we include contributions to the Zeeman energy 
that are diagonal in the flux sector, with local fields $\hh_{1}$ and $\hh_{2}$ acting near the two vacancies; when vacancy separation is large, we find again that only
the $z$-components enter the energy at leading order, and we denote these by $h_1$ and $h_2$.

We will show, extending Eq.~(\ref{vac-fermion}), that $H_u$ for the two-vacancy problem at weak field, projected onto low energy states, has the form
\begin{eqnarray}\label{2vacGappedEffHam}
H_u &=& \n h_1 (2a^\dagger_1a_1 -1) + \n h_2 (2a^\dagger_2a_2 -1) \nonumber \\
&&+\, {\rm i} \varepsilon(a^\dagger_1 + a_1)(a^\dagger_2 + a_2)
\,.
\end{eqnarray}
As for a single vacancy, in the gapped phase with $0\leq j_{x},j_{y}<1$, only the $z$-components of local moments develop non-zero values in weak fields. We can write spin operators at sites adjacent to each of the vacancies within a flux sector in terms of complex fermions. Projecting onto the four-dimensional space of low energy states, and denoting the projection operator by ${\cal Q}$, we find 
${\cal Q}\tilde{\sigma}^z_{m}{\cal Q}= \n (2a_m^\dagger a_m -1)$, with $m=1$ or $2$ labeling the vacancies.


We next derive the effective Hamiltonian (\ref{2vacGappedEffHam}) and calculate the energy $\varepsilon$.
We do this by using a T-matrix approach to relate the Green function for the tight binding model
that represents the system with vacancies to the Green function for the undiluted hexagonal lattice.
To establish some notation, consider matrices $H$, $H_0$ and $V$, related by 
$H = H_{0}+V$, and define the Green functions $G=(z-H)^{-1}$ and $G_0=(z-H_0)^{-1}$.
Then the T-matrix is 
\begin{equation}\label{t-matrix-def}
T=V(1-G_{0}V)^{-1}
\end{equation}
and 
\begin{equation}\label{t-matrix}
G =G_{0}+G_{0}TG_{0}\,.
\end{equation}
In general, if the spectrum of $H_0$ has an energy gap and $H$ has levels within this gap, $T$ has poles in the complex $z$-plane at the locations of these levels. We determine $\varepsilon$ by finding these poles.

For clarity it is convenient to relate the undiluted lattice to the system with vacancies in two steps:
in the first step we eliminate site orbitals at the locations of the vacancies; in the second we couple
three additional orbitals around each vacancy, as illustrated in Fig.~\ref{EffHam}(b).
To implement the first step we choose $V$ to be  a potential 
$1/\epsilon$ acting at each of the vacancy sites, and take the limit $\epsilon\rightarrow 0$.

Consider a B vacancy in the unit cell at the origin and an A vacancy in the unit cell at $\rr$, 
with $\rr = n_1 {\bf n}_1 + n_2 {\bf n}_2$.  
Let $G^{\alpha\beta}_{F}(\rr_1,\rr_2)$ be the Green function for a lattice without vacancies but with fluxes through one of the plaquettes adjacent to each of the sites where vacancies will be introduced. The poles of $T$ are at the values of $z$ for which
\begin{equation}
G^{AA}_{F}(\rr,\rr)G^{BB}_{F}(\mathbf{0},\mathbf{0})-G^{AB}_{F}(\rr,\mathbf{0})G^{BA}_{F}(\mathbf{0},\rr)=0\,.
\end{equation}

We are unfortunately able to compute $G^{\alpha\beta}_{F}(\rr_1,\rr_2)$ only numerically.
We find however (in the sense made precise below) that key features of its behaviour are the same as for the lattice without fluxes. In the zero flux sector convenient analytical expressions 
are available for the Green function, which we denote by $G^{\alpha\beta}_{0}(\mathbf{0},\rr)$,
and we base our initial discussion on these. We will see that $\varepsilon$ decreases exponentially with vacancy separation. To study well-separated vacancies we therefore require the Green function at $z$ small compared to the gap $\Delta$. An expansion in powers of $z/\Delta$ gives 
(see Appendix \ref{GrFnHexLattice}) 
\begin{equation}\label{GAAandGBA}
\begin{array}{rl}
G^{AA}_{0}(\mathbf{0},\mathbf{0}) = & \dfrac{-z}{J_{z}^{2}\sqrt{(1-j_{x}^{2}-j_{y}^{2})^{2}-4j_{x}^{2}j_{y}^{2}}}+O\left[\frac{1}{J_{z}}\left(\frac{z}{\Delta}\right)^{3}\right]\\[6ex]
G^{BA}_{0}(\mathbf{0},\rr) = &\dfrac{(-1)^{n_{1}+n_{2}+1}}{J_{z}}j_{x}^{n_{1}}j_{y}^{n_{2}}{n_{1}+n_{2} \choose n_{1}}+O\left[\frac{1}{J_{z}}\left(\frac{z}{\Delta}\right)^{2}\right] \\[4ex]
&\qquad n_{1}, n_{2} \geq 0,\text{ 0 otherwise}
\end{array}
\end{equation}
with the other Green function elements obtained by symmetry. 

From this we find that two vacancies on opposite sublattices, located so that their zero-mode wedges overlap produce a pair of levels within the gap, with energies of magnitude
\begin{equation}
\varepsilon = {\cal N}^2 J_z j_x^{n_1} j_y^{n_2} {n_{1}+n_{2} \choose n_{1}}
\end{equation}
when $\rr$ is large compared to the decay length of the Green function.
We also find, from a numerical study, that in the gapped phase $G_F^{AA}(\rr,\rr)/G_0^{AA}(\rr,\rr)$ and  $G_F^{BA}(\mathbf{0},\rr)/G_0^{AA}(\mathbf{0},\rr)$ approach unity when $|\rr|$ is large compared to the decay length of the Green function. Behaviour of well-separated vacancy pairs is therefore the same in both the ground-state and the flux-free sectors.

\begin{figure}
\epsfig{file=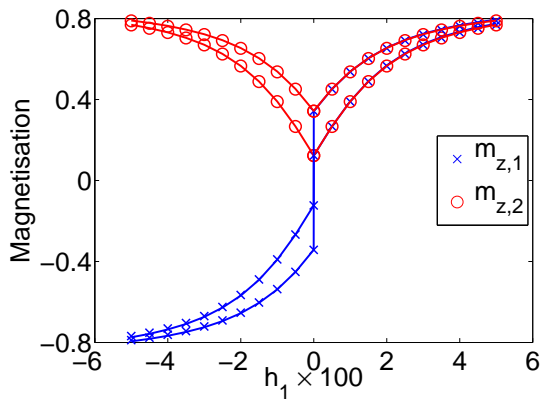,width=0.4\textwidth}
\caption{\label{EffSpinHamFig} Local moments formed near each vacancy, as a function of field strength, in a system with two vacancies on opposite sublattices:  $m_{z}$ vs. $h_{1}$ for (from the top) $h_{2} = \varepsilon/2$ and $h_{2} = \varepsilon/6$.
Interaction strengths $j_x=j_y=1/3$; vacancy separation $10({\bf n}_1 + {\bf n}_2)$; mode splitting $\varepsilon = 7.9 \times 10^{-3} J_z$; saturation moment $\n  = 0.863$. Points are obtained from the projected Hamiltonian, Eq.~(\ref{2vacGappedEffHam}); curves are 
for the full Hamiltonian in the flux-free  sector using lattice of size $20\times 20$.}
\end{figure}

In a second step we include the basis orbitals in the tight binding model that arise from unpaired 
Majorana fermions $b_0^\alpha$ and $b_{\rr}^\alpha$. 
As for a single vacancy, the leading contributions arise only from $\alpha = z$ when $|\rr|$ is large.
We are therefore concerned with four states within the gap, and provided Zeeman fields are sufficiently small we can project onto this subspace. Doing so, we obtain a tight binding model of the form
\begin{equation}
\left(
\begin{array}{cc}
0 & M\\
M^T & 0
\end{array}
\right) \quad {\rm with} \quad
M = 
\left(
\begin{array}{cc}
\n h_1 & 0\\
\varepsilon & \n h_2
\end{array}
\right)\,.
\end{equation}
The Majorana fermion Hamiltonian
\begin{equation}
H_{u} ={\rm i}\left(
\begin{array}{cc}
b_A & c_A
\end{array}\right)
M
\left(
\begin{array}{c}
c_B \\ b_B
\end{array}\right)
\end{equation}
can readily be rewritten in the form of Eq.~(\ref{2vacGappedEffHam}). 
It has eigenvalues $\pm S_{+}$ and $\pm S_{-}$ given by
\begin{equation}
\begin{array}{ll}
S_{+} &= \frac{1}{2}\left[\sqrt{\varepsilon^{2}+{\cal N}^2(h_{1}+h_{2})^{2}}+\sqrt{\varepsilon^{2}+{\cal N}^2(h_{1}-h_{2})^{2}}\right]\\[2ex]
S_{-} &= \frac{1}{2}\left|\sqrt{\varepsilon^{2}+{\cal N}^2(h_{1}+h_{2})^{2}}-\sqrt{\varepsilon^{2}+{\cal N}^2(h_{1}-h_{2})^{2}}\right|\nonumber
\end{array}
\end{equation}
and so Eq.~(\ref{2vacGappedEffHam}) has the form
\begin{equation}
H_u =S_{+}(2a_{+}^{\dagger}a_{+}-1)+S_{-}(2a_{-}^{\dagger}a_{-}-1)\,.
\end{equation}
For example, with $h_{1}=h_{2}=h$, this gives a moment for the system as a whole, of
\begin{equation}\label{mag2vacGapped}
m(\varepsilon,h) = \dfrac{4  \n^{2}h}{\sqrt{\varepsilon^2+4h^{2}\n^{2}}}\,.
\end{equation}
For fields $h\ll \varepsilon$ there is a large, field independent impurity magnetic susceptibility $\chi = 4\n^{2}/ \varepsilon$, and for $\varepsilon\ll h$ the induced moment saturates at $2\n $, twice the isolated vacancy moment.
In Fig.~\ref{EffSpinHamFig} we illustrate behaviour for $h_1\not= h_2$. This figure also demonstrates that projection onto states within the gap provides a very accurate
treatment of the full Hamiltonian with Zeeman energy of the form given in Eq.~(\ref{ProjZeemanEn}).
As for a single vacancy, from this form of the Zeeman energy it is apparent that the moment is entirely localised on the sites equivalent to $\rr_{3}$ in Fig.~\ref{EffHam}(b), adjacent to each vacancy.

\subsection{Finite density of vacancies}\label{finite density}
We next investigate the properties of the Kitaev model in the gapped phase with a finite density of vacancies. We will see that couplings between the low-energy degrees of freedom generate an impurity band with a density of states that has a Dyson-type divergence at zero energy. This leads to a divergent \textit{macroscopic} susceptibility.

The approach to the two vacancy problem outlined above can be extended to include a finite density of vacancies. Consider a system with $N_A$ vacancies on the $A$-sublattice and $N_B$ vacancies on the $B$-sublattice, and let $G_{F}^{AA}$, $G_{F}^{BB}$ and $G_{F}^{AB}$ be matrices of Green function elements for the undiluted lattice (all functions of $z$) evaluated between the vacancy sites. The energy levels of the honeycomb lattice tight binding model with these vacancies are given by the values of $z$  for which
\begin{equation}
\left|\begin{array}{cc}G_{F}^{AA}&(G_{F}^{BA})^{\text{T}}\\[1ex]
G_{F}^{BA}&G_{F}^{BB}\end{array}\right| = 0.
\end{equation}

We assume that vacancy separations are much larger than the decay length of the Green function. Then $N_A + N_B$ vacancies give rise to an impurity band of $N_A + N_B$ levels, with energy width $\ll \Delta$, at the centre of the band gap of the undiluted lattice. In these circumstances we are concerned with $|z|\ll \Delta$ and the Green function elements can be approximated by the leading order in an expansion in $z/\Delta$. At this order $G_F^{AB}$ is evaluated at $z=0$ and 
\begin{equation}
G^{AA}_F  \approx z[\partial_z G_F({\bf 0},{\bf 0})|_{z=0}] \cdot \openone = -(z/{\cal N}^2J_z^2)\cdot \openone\nonumber \,,
\end{equation}
and similarly for $G^{BB}_F$.
Within these approximations, the impurity band levels are therefore the eigenvalues of the bipartite hopping Hamiltonian
\begin{equation}
H_{\text{BRH}} = 
{\cal N}^2 J_z^2 \left(\begin{array}{cc}0&(G_{F}^{BA})^{T}\\
G_{F}^{BA}&0\end{array}\right).
\label{BRH}
\end{equation}
Our focus is on the case of compensated vacancies ($N_A=N_B$) for which the impurity band is generically free of zero modes; by contrast uncompensated vacancies result in at least $|N_A-N_B|$ zero modes.


The Hamiltonian of Eq. (\ref{BRH}) is one example from the well-studied class of bipartite random hopping (BRH) models. Such models are characterised by a density of states $\rho(E)$ that is strongly divergent as energy $E$ approaches zero, having the form $\rho(E)= \mathcal{F}(E)/|E|$, where $\mathcal{F}(E)$ is a function that goes to zero more slowly than any power of $E$ but ensures that the density of states is integrable.\cite{Gade:1993,Damle2002} While renormalisation group treatments of these models typically generate broadly distributed coupling strengths, even our bare Hamiltonian already has couplings that vary over many orders of magnitude since they depend exponentially on vacancy separation.

\begin{figure}[h]
\epsfig{file=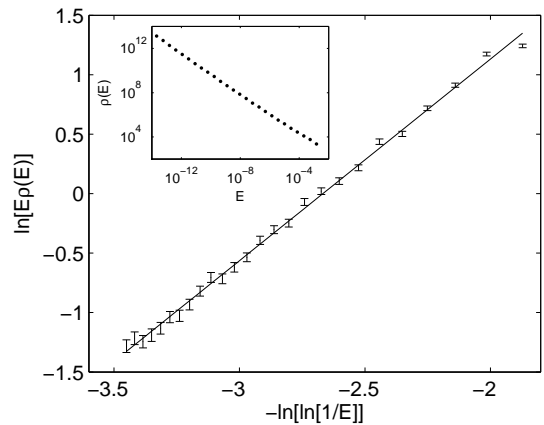,width=0.4\textwidth}
\caption{Density of states for 160 randomly placed compensated vacancies on a lattice of 3200 sites, with fluxes attached to vacancies and with $J_{x} = J_{y} = 1$, $J_{z}=4$.  The density of states is an average over 1000 disorder realisations. Solid line is a best fit to the data with gradient of $1.7$. 
Inset: The same data, plotted on a scale that illustrates the very large energy range considered.}
\label{DoS}
\end{figure}

We find from a numerical study of Eq.~(\ref{BRH}) a strong divergence in the density of states, over about ten decades,
as illustrated in  Fig.~\ref{DoS}.
While the exact behaviour of $\mathcal{F}(E)$ is difficult to ascertain, our results are compatible with the form
\begin{equation}
\mathcal{F}(E)\propto \dfrac{1}{\log[1/E]^{x}}~.
\end{equation}
and a fit yields $x = 1.7$.

We next examine response to a Zeeman field. Consider first the impurity band Hamiltonian in the
absence of a field. The off-diagonal blocks of $H_{\rm BRH}$ have a singular value decomposition
$G_F^{BA} = vsu^{\rm T}$, where (taking $N_A=N_B$) $v$ and $u$ are $N_A\times N_A$ orthogonal matrices and $s$ is an $N_A\times N_A$ diagonal matrix with positive entries $s_j$. The impurity band Hamiltonian is therefore reduced to a direct sum of $2\times 2$ block-diagonal forms 
by the transformation $W^{\rm T}H_{\rm BRH}W$, with
\begin{equation}
W = \left(\begin{array}{cc}
v&0\\
0&u
\end{array}\right)\,.
\end{equation}
Coupling to the projected Zeeman field [Eq.~(\ref{ProjZeemanEn})] is invariant under this transformation, provided the field strengths $h_\alpha$ acting near each vacancy are the same. For this reason,
each pair of levels from the impurity band with energies $\pm s_j$ makes a contribution to the Zeeman response like that from a single pair of vacancies with $\varepsilon = s_j$. Following Eq. (\ref{mag2vacGapped}),  
the total magnetisation and susceptibility are then 
\begin{equation}
\begin{array}{rl}
m=& \dint m(E,h)\rho(E) dE\,,\\
\chi = & \dint\dfrac{4\n^{2}E}{(E^{2}+4\n^{2}h^{2})^{3/2}}\mathcal{F}(E)dE\,.
\end{array}
\end{equation}
The susceptibility integral has its largest contribution from a region around $\sqrt{2}\n  h$. Since $\mathcal{F}(E)$ 
is slowly varying,
the susceptibility has the form
\begin{equation}\label{SusceptibilityMultiVacGap}
\chi\simeq 
\frac{\mathcal{F}\left(\sqrt{2}\n  h\right)}{h}\,.
\end{equation}
This singular behaviour arises because well-separated vacancy pairs give rise to local moments
that are fully polarised even in a weak field.
\section{Gapless Phase}
\label{GaplessPhase}
The response of the Kitaev model with vacancies to a Zeeman field is very different in the gapless phase compared to that in the gapped phase. The difference arises because the impurity modes that form a finite-dimensional low-energy subspace in the gapped phase become continuum resonances in the gapless phase. As a physical consequence, magnetic response in the gapless phase
has striking singularities at weak field, which we discuss in this section. 

Behaviour is qualitatively the same throughout the phase, and we focus on the isotropic 
point $J_x{=}J_y{=}J_z{\equiv}J$. We will be concerned with the magnetisation of a system
with a single vacancy in a Zeeman field that is weak compared to $J$, and with
properties of a pair of vacancies that have a separation large compared to the lattice spacing. 
A feature of the gapless phases is that the zero mode induced by a single vacancy without Zeeman coupling is merely power-law localised in the gapless phase, and indeed is not normalisable in an infinite system.\cite{pereira-2006-96} 
Moreover, its probability density is not confined to a wedge (as in the gapped phase), but instead approximately
isotropic. One direct consequence is that there is a large response to all components
of the Zeeman field, rather than just $h_z$ as is the case for $0\leq j_x,j_y<1$.

We first comment on the ground state flux sector in a system with a vacancy. The approach used
above (see Section \ref{flux binding}) to discuss flux binding to a vacancy deep in the gapped phase
is not useful in the gapless phase because the expansion parameters $j_x, j_y$ are not small.
Instead, as we have reported elsewhere, \cite{willans2010} a direct numerical calculation can be used
to show that a vacancy binds a flux with energy $-0.027J$.

\subsection{Magnetisation of a single vacancy}
In this subsection we calculate the magnetic response of the Kitaev model
with a single vacancy in the gapless phase. We present results for both the ground state
flux sector, with a flux bound to the vacancy, and for the flux-free sector. 
The second case has the advantage that calculations are simpler. In addition, it turns out to
be relevant to behaviour of a system with two vacancies that both have bound fluxes: since we
are dealing with $Z_2$ fluxes, low-energy properties in a system with two nearby fluxes
are like those in the zero flux sector.

A summary of the main results is as follows. In the ground state flux sector the vacancy susceptibility
at weak field $h$ and low temperature $T$ diverges as 
\begin{equation}
\chi\propto\left\{\begin{array}{l l}\label{chi-non-zero-flux}
\ln(1/h) & \quad {\rm for} \quad T\ll h\\
\ln(1/T) & \quad {\rm for} \quad h\ll T\,.
\end{array}\right. 
\end{equation}
In the zero flux sector, the vacancy susceptibility has the still stronger singularities
\begin{equation}\label{chi-zero-flux}
\chi\propto\left\{\begin{array}{l l}
1/[h ( \ln(1/h) )^{3/2} ]& \quad {\rm for}\quad T\ll h/\sqrt{\ln(1/h)} \\
1/[T \ln(1/T)]
& \quad {\rm for} \quad  h/\sqrt{\ln(1/h)}\ll T\,.
\end{array}\right. 
\end{equation}

To derive these results, we require the ground state energy or (at finite $T$) free energy of the 
fermionic degrees of freedom in the relevant flux sector. These follow from the eigenvalues
of the tight-binding Hamiltonian, and the necessary information is contained in the trace of the
Green function and its dependence on the complex energy $z$. We express this Green function
in terms of the one for a system without vacancies using the $T$-matrix approach described
above [see Eq.~(\ref{t-matrix-def})]. 

Let  $G(\hh,\rr,\rr')$ be the matrix element between sites $\rr$ and $\rr'$ of the Green function for the tight binding model of Fig. \ref{EffHam}(b). We define the difference between the Green function trace at finite field $\hh$ and at zero field, as
\begin{equation}\label{rho-def}
\rho(z,h) = \text{Tr}[G(\hh,\rr,\rr')-G(0,\rr,\rr')]\,.
\end{equation}
The discontinuity in the imaginary part of $\rho(z,h)$ across the real $z$ axis gives in the usual way the difference between the finite and zero field density of states for the tight binding model, and so
the ground state energy difference is expressed by the integral
\begin{equation}\label{energy-integral}
\mathcal{E}(h) = \dfrac{1}{2\pi i}\oint z\rho(z,h)dz\,,
\end{equation}
taken in an anti-clockwise direction on a contour enclosing the negative real axis.
Similarly, the free energy difference, computed within the flux sector, is
\begin{equation}\label{free-energy-integral}
\mathcal{F}(h) = -\dfrac{1}{2\pi i}\doint T\rho(z,h)\ln\left[2\cosh(z/T)\right]dz\,,
\end{equation}
evaluated on the same contour.

We next outline the evaluation of ${\cal E}(h)$, taking $\hh = (0,0,h)$
for simplicity of presentation. In this case we can omit the orbitals arising from
the Majorana fermions $b^x$ and $b^y$, retaining only the one from $b^z$ (see Fig.~\ref{EffHam}).
We start with a lattice of $2N$ sites before dilution 
and give the $b^z$-orbital the site label $\rr_z$. 
%
%
To obtain the Green function $G(0,\rr,\rr')$ for a system with a vacancy,
we introduce a potential $1/\epsilon$ at the vacancy site $\rr_{v}$ and take the limit $\epsilon\rightarrow 0$. 
In this limit $G(0,\rr,\rr')$ 
consists of one $(2N-1)\times (2N-1)$ block, with elements
\begin{equation}
G_{0}(\rr,\rr')-\dfrac{G_{0}(\rr,\rr_{v})G_{0}(\rr_{v},\rr')}{G_{0}(\rr_v,\rr_v)} 
\end{equation}
for $\rr,\rr' \neq \rr_{v}, \rr_{z}$ and one $1\times 1$ block with element $1/z$ for the site $\rr_z$. 
We include non-zero $h$ by applying the $T$-matrix approach a second time, with
$G(0,\rr,\rr')$ as the initial Green function, taking
\begin{equation}
V = \left(\begin{array}{cc}
0& h\\
h & 0
\end{array}\right)
\end{equation}
in the basis of sites $\rr_z, \rr_3$ (see Fig.~\ref{EffHam}).
With the shorthand
$g(z) \equiv G(0,\rr_{3},\rr_{3}) $, 
the $T$-matrix is 
\begin{equation}
T = \dfrac{h}{z-g(z)h^{2}}\left(\begin{array}{cc}g(z)hz&z\\z&h\end{array}\right).
\end{equation}
Finally, using $\sum_{r} G(0,\rr,\rr_{3})G(0,\rr_{3},\rr) = -\partial_{z}g(z)$, we obtain
\begin{equation}\label{rho-z-h}
\rho(z,h) = h^2[z^{-1}g(z)-\partial_zg(z)]/[z-h^2g(z)]\,.
\end{equation}
Setting $g(z) - z \partial_z g(z) = a(z) + {\rm i} b(z)$ and $z-h^2g(z) = u(z) + {\rm i} v(z)$, with $a(z)$, $b(z)$, $u(z)$ and $v(z)$ real, Eq.~(\ref{energy-integral}) becomes
\begin{equation}\label{E-new}
{\cal E}(h) = \frac{h^2}{\pi} \int_{-\infty}^0 {\bf d}x \, \frac{b(x)u(x) - a(x)v(x)}{u^2(x)+v^2(x)}\,.
\end{equation}

To make use of these results, we require matrix elements of the Green function for
the undiluted lattice. At small $h$, the function $u(x)$ has a zero at $x=x_0$, with $-1\ll x_0<0$, and the dominant contribution to the integral in Eq.~(\ref{E-new}) comes from the vicinity of this point. We are therefore concerned with the
Green function at small $z$: it has the
behaviour (see Appendix \ref{GrFnHexLattice})
\begin{equation}\label{GreensFn}
G_{0}(\rr,\rr) \sim \lambda z\ln\left[-(\mu z)^2\right] \quad
G_{0}(\rr_{3},\rr_{v}) \sim -\nu~,
\end{equation}
where, for $J_{\alpha}=1$, $\lambda = 1/\sqrt{3} \pi$ and $\mu = \nu = 1/3$.  The $z$-dependences of Eq. (\ref{GreensFn}) are a direct consequence of the massless Dirac spectrum for the nearest neighbour tight binding model on the honeycomb lattice, and hold with appropriate values for $\lambda$, $\mu$ and $\nu$ throughout the gapless phase.

We expand the integrand of Eq.~(\ref{E-new}) about $x=x_0$ in the small-$h$ limit, retaining only the leading terms. 
Using the asymptotic forms for Green function elements given in Eq.~(\ref{GreensFn}), we find
\begin{equation}
u(x) \approx x + \frac{h^2 \nu^2}{2\lambda x \ln x}
\end{equation} 
and therefore
\begin{equation}
x_0 \approx -\frac{h\nu}{\sqrt{2\lambda \ln(1/h)}}\,.
\end{equation}
Then $\partial_x u(x)|_{x=x_0} \approx 2$, and writing $x=x_0 + s$ we have
\begin{equation}
{\cal E}(h) \approx \frac{h^2}{\pi} \int_{-\infty}^\infty {\rm d} s\, \frac{a(x_0)v(x_0)}{4s^2 + v^2(x_0)} = \frac{h^2  a(x_0)}{2}\,.
\end{equation}
Moreover, $h^2 a(x_0) \approx 2x_0$,  which yields the energy
\begin{equation}\label{energy-result}
\mathcal{E}(h) \sim -\frac{h \nu}{\sqrt{2\lambda\ln(1/h)}}\,,
\end{equation} 
the magnetisation 
\begin{equation}
m(h)= -\partial_{h}\mathcal{E}(h)\sim \frac{\nu}{\sqrt{2\lambda \ln(1/h)}}
\label{EandM}
\end{equation}
and the susceptibility, Eq.~(\ref{chi-zero-flux}). Results for non-zero temperature are obtained in a similar way.

Behaviour for a general field orientation can be obtained from a similar, although more involved, calculation.
We find that, even in the general case, only the field magnitude enters the leading contribution to $\rho(z,h)$
at small $h$, which is therefore orientation-independent.
As for the gapped phase, the magnetisation is entirely localized on sites adjacent to the vacancy but now each of these sites, labelled $\rr_{1}, \rr_{2}, \rr_{3}$ in Fig. \ref{EffHam}(b), carries a separate component ($m_{x}, m_{y}$ and $m_{z}$, respectively), 
proportional to the corresponding component of $\hh$: as the field orientation changes, the induced magnetisation moves around the vacancy in real space!  


\begin{figure}
\begin{tabular}{c}
\epsfig{file=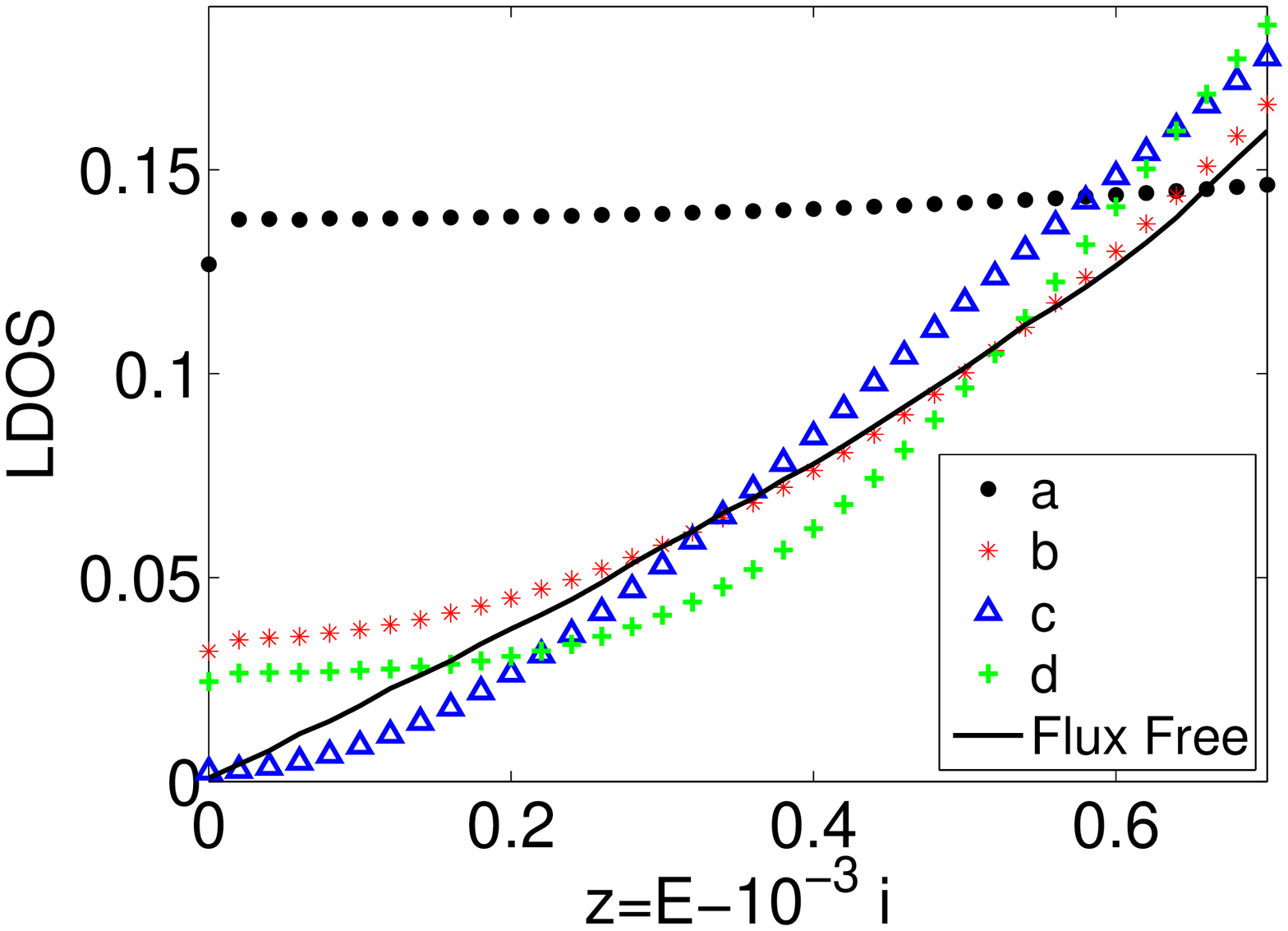,width=6cm}\quad\epsfig{file=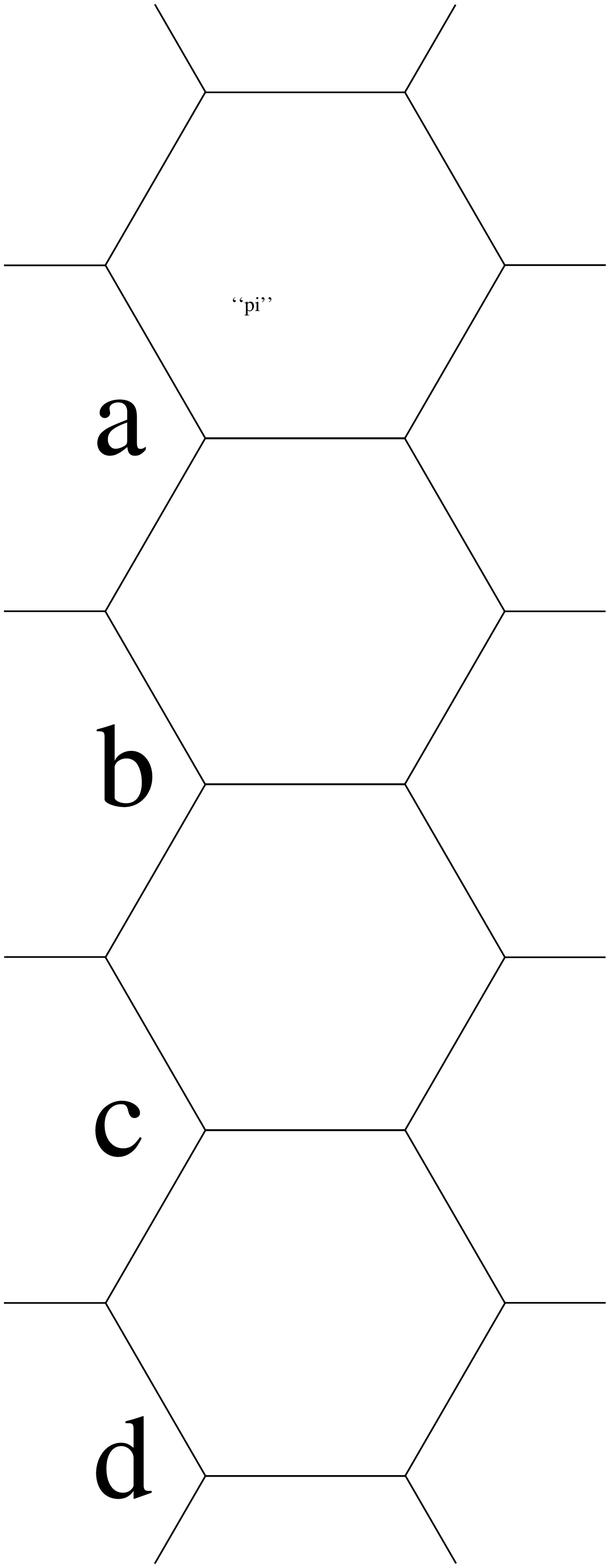,width=2cm}\\[2ex]
\epsfig{file=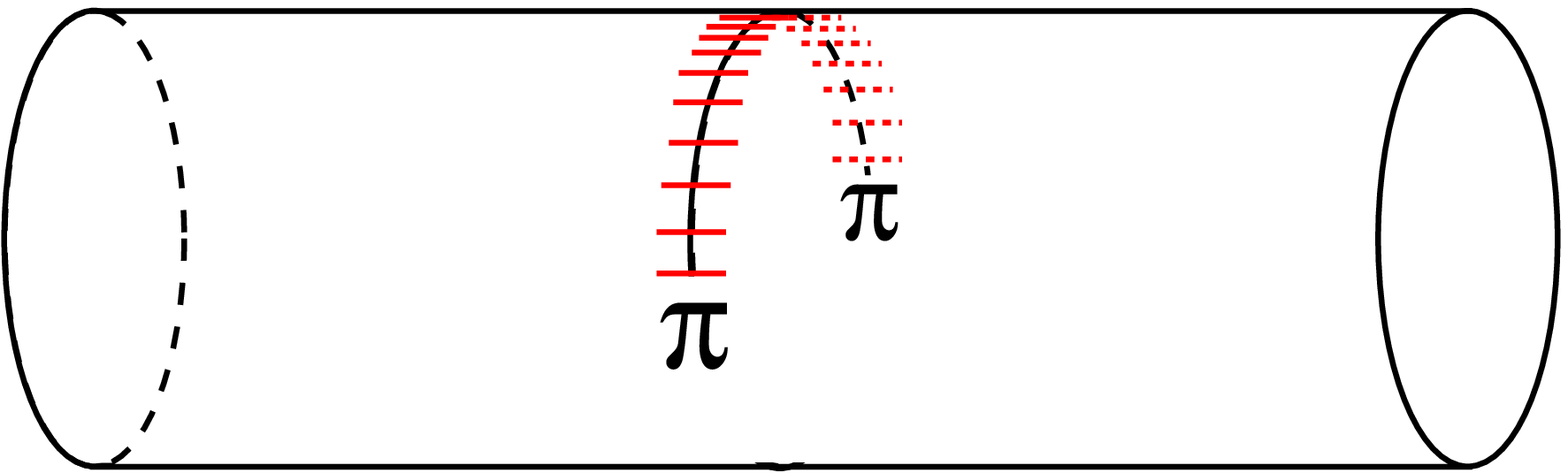,width=7cm}
\end{tabular}
\caption{Results from a numerical study of the effect of a flux on the local density of states (LDOS). 
{\it Bottom}: geometry of the system, showing two fluxes (indicated by $\pi$'s), joined by a string. The sign of the hopping matrix element on the links crossed by the string is reversed in the presence of the fluxes. {\it Top Right}: detailed view of the geometry, showing one flux ($\pi$) through a plaquette, and the labeling ($a$, $b$, $c$ and $d$) of nearby sites.
{\it Top Left}: LDOS as a function of energy $E$. Points: with a flux, at the four labeled sites as indicated; line: without flux (same behaviour at all sites).}
\label{LDOSflux}
\end{figure}

We next turn to behaviour in the ground state flux sector, with a flux attached to the vacancy plaquette.
All relevant information is contained in the function $g(z)$, and in particular, its form for small $z$.
This in turn depends on site-diagonal and nearest-neighbour Green function elements. Moreover, since
the nearest-neighbour Green function elements are finite at $z=0$ in both flux sectors, the crucial quantity
is the site-diagonal element, or equivalently the local density of states (LDOS). In the zero flux sector this varies linearly
at small energy, reflecting the Dirac cones in the honeycomb tight binding model dispersion. 
We use a numerical study to find the LDOS at small energy in the presence of a flux. The geometry employed and the results obtained are illustrated in Fig.~\ref{LDOSflux}.
We evaluate the LDOS at sites close to a plaquette threaded by an isolated flux,
for a cylindrical system of infinite length and finite circumference (taken to be 1502 lattice units for the data shown). 
We find that the flux has a dramatic effect, resulting in a finite LDOS in its vicinity at small energy.
In consequence, whereas $g(z)$ is divergent for $z\to 0$ in the zero flux sector, it has a finite limit 
in the ground state sector. The resulting weak-field form for the susceptibility is shown in Eq.~(\ref{chi-non-zero-flux}).


\subsection{Zeeman response with two vacancies}

The response to a Zeeman field of a system in the gapless phase with a pair of vacancies involves the physics of
the Kitaev model in particularly rich ways. In summary, we find three types of behaviour in the ground state flux sector, depending on the field strength $h$, the vacancy separation, and whether vacancies lie on the same or opposite sublattices. The magnitude of the vacancy separation vector $\dd$ sets a field scale $h_{\rm c}$, which decreases with increasing separation. $(i)$ For {\em fields large on this scale} 
the response of a pair of vacancies is simply the sum of the responses that would arise for each in isolation. 
By contrast, weak field behaviour depends on the relative sublattices of the vacancies. $(ii)$ {\em Two vacancies on
opposite sublattices} generate a contribution to the susceptibility that is field-independent at $h \ll h_{\rm c}$
and parametrically larger at large $|\dd|$ than the susceptibility per spin of the host system. $(iii)$ Most strikingly of
all, {\em two vacancies on the same sublattice} have a susceptibility that is parametrically more strongly divergent
for $h\ll h_{\rm c}$ than for an isolated vacancy, being of the same form as for a single vacancy in the zero flux sector,
as given in Eq.~(\ref{chi-zero-flux}). In essence, this is because two vacancies bind a pair of $Z_2$ fluxes
in the ground state sector, and because in their influence on low-energy properties these fluxes effectively fuse  and cancel. 

We derive these results by expressing the energy ${\cal E}(h)$ of the system
with vacancies and a field in terms of Green function elements for the
undiluted lattice, using extensions of the T-matrix methods outlined above. In the zero flux sector, analytical 
expressions for these elements are available, which we supplement with computational results in the
ground state flux sector. Since calculations are quite involved, we omit many details and consider only the case of projected Zeeman fields
[Eq.~(\ref{EffHam})] oriented along the $z$-axis and of equal strength $h$ for both vacancies. 

\begin{figure}[h]
\epsfig{file=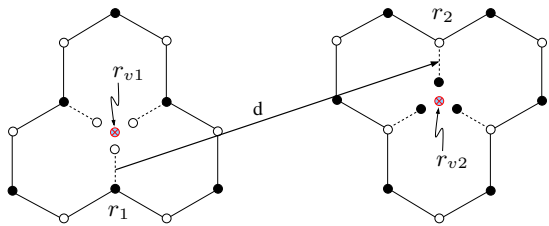,width=0.4\textwidth}
\caption{\label{vacABfig}Site labelling used in our calculation of the Zeeman response of a system with  two vacancies on opposite sublattices. Similar notation is used for vacancies on the same sublattice, where $\rr_{1}$ and $\rr_{2}$ are the sites adjacent to the vacancies in the $z$-direction, selected by our choice of field orientation.}
\end{figure}

As in our discussion of a system with a single vacancy in the gapless phase, it is convenient to
separate calculations into two steps, introducing vacancies at the first step, and including a Zeeman field
at the second. Some notation is summarised in Fig.~\ref{vacABfig}. The Green function $G(0,\rr,\rr')$ describing a
system with $N$ unit cells after the first step is a block diagonal matrix consisting of 
two $1\times 1$ blocks with entries $1/z$, from the $b^z_j$ orbitals, and
one $(2N-2)\times (2N-2)$ block
with entries
\begin{eqnarray}
G(0,\rr,\rr') &=& G_0(\rr,\rr')\nonumber\\
&+& (G_0(\rr,\rr_{v1}) G_0(\rr,\rr_{v2})) T_v \left(\begin{array}{c}
G_0(\rr_{v1},\rr')\\
G_0(\rr_{v2},\rr') 
\end{array}\right) \nonumber
\end{eqnarray}
for $\rr,\rr'\not= \rr_{v1}\,, \rr_{v2}, \, \rr_{z1}\,{\rm or}\,\, \rr_{z2}$, the form of the $T$-matrix being
\begin{equation}
T_v = -\left(\begin{array}{cc}
G_0(\rr_{v1},\rr_{v1})& G_0(\rr_{v1},\rr_{v2})\\
G_0(\rr_{v2},\rr_{v1})& G_0(\rr_{v2},\rr_{v2})
\end{array}\right)^{-1}\,.
\end{equation}
From this, using the definition given in Eq.~(\ref{rho-def}), we find
\begin{equation}\label{rho2}
\rho(z,h) = N(z,h)/D(z,h)
\end{equation}
with (after a lengthy calculation)
\begin{eqnarray}\label{numerator}
N(z,h) &=& 2h^2\{z^{-1}h^2[G(0,\rr_1,\rr_2)^2 \nonumber\\
&&- G(0,\rr_1,\rr_1)^2] +G(0,\rr_1,\rr_1)\}\nonumber\\
&-&2h^2[z-h^2G(0,\rr_1,\rr_1)]\partial_zG(0,\rr_1,\rr_1) \nonumber\\
&-& 2h^4 G(0,\rr_1,\rr_2) \partial_z G(0,\rr_1,\rr_2)\nonumber 
\end{eqnarray}
and
\begin{equation}\label{denominator}
D(z,h) = [z-h^2 G(0,\rr_1,\rr_1)]^2 - h^4 G(0,\rr_1,\rr_2)^2\nonumber
\end{equation}
where we have used the symmetries of $G(0,\rr,\rr')$ to simplify expressions.

Behaviour in each of the cases $(i)$--$(iii)$ summarised above can be extracted by considering simplifications of
Eq.~(\ref{rho2}) in the relevant limits. First we note that the contributions to $N(z,h)$ and $D(z,h)$ involving $G(0,\rr_1,\rr_2)$ may be omitted when discussing weak field behaviour, because they appear only
in terms that are ${\cal O}(h^4)$. With this simplification, $\rho(z,h)$ reduces to twice the expression that applies in a system with a single vacancy [as given in Eq.~(\ref{rho-z-h})] but with $G(0,\rr_1,\rr_1)$ evaluated for the two vacancy system. We therefore need to examine the behaviour of $G(0,\rr_1,\rr_1)\equiv g(z)$  in this case. The integrals in Eqns.~(\ref{energy-integral}) and (\ref{free-energy-integral}) are dominated by contributions from $|z| \sim {\cal O}(h)$ (omitting for simplicity logarithmic corrections), and so we are concerned with $g(z)$ at the scale $|z|\sim h$. We first discuss this and its consequences in the zero flux sector.

$(i)$ {\it Independent vacancies.}  Coupling between vacancies in the expressions for $G(0,\rr,\rr')$ involves $G_0(\rr_{v1},\rr_{v2})$. This is exponentially small in $|\dd|$ unless $|z| \lesssim |\dd|^{-1}$, which leads us to identify $h_{\rm c} = J/|\dd|$: for $h\gg h_{c}$ we can neglect $G_0(\rr_{v1},\rr_{v2})$ and the response is a sum of independent contributions arising from each of the two vacancies.
$(ii)$ {\it Coupled vacancies on opposite sublattices.} For $h\ll h_{\rm c}$ and compensated vacancies, we find that $g(z) \sim |\dd|^2 z$ if $|z| \ll |\dd|^{-1}$ and is small otherwise. In consequence, we obtain a field-independent susceptibility that is of order $|\dd|$ and hence much larger than the susceptibility per spin of the undiluted system.
$(iii)$ {\it Coupled vacancies on the same sublattice.} In this case, using the forms for the Green function of the undiluted lattice, discussed in Appendix B, we find the same singularity in $g(z)$ at small $z$ as for a single vacancy, and therefore obtain the same singularity in the susceptibility.

\begin{figure}[t]
\epsfig{file=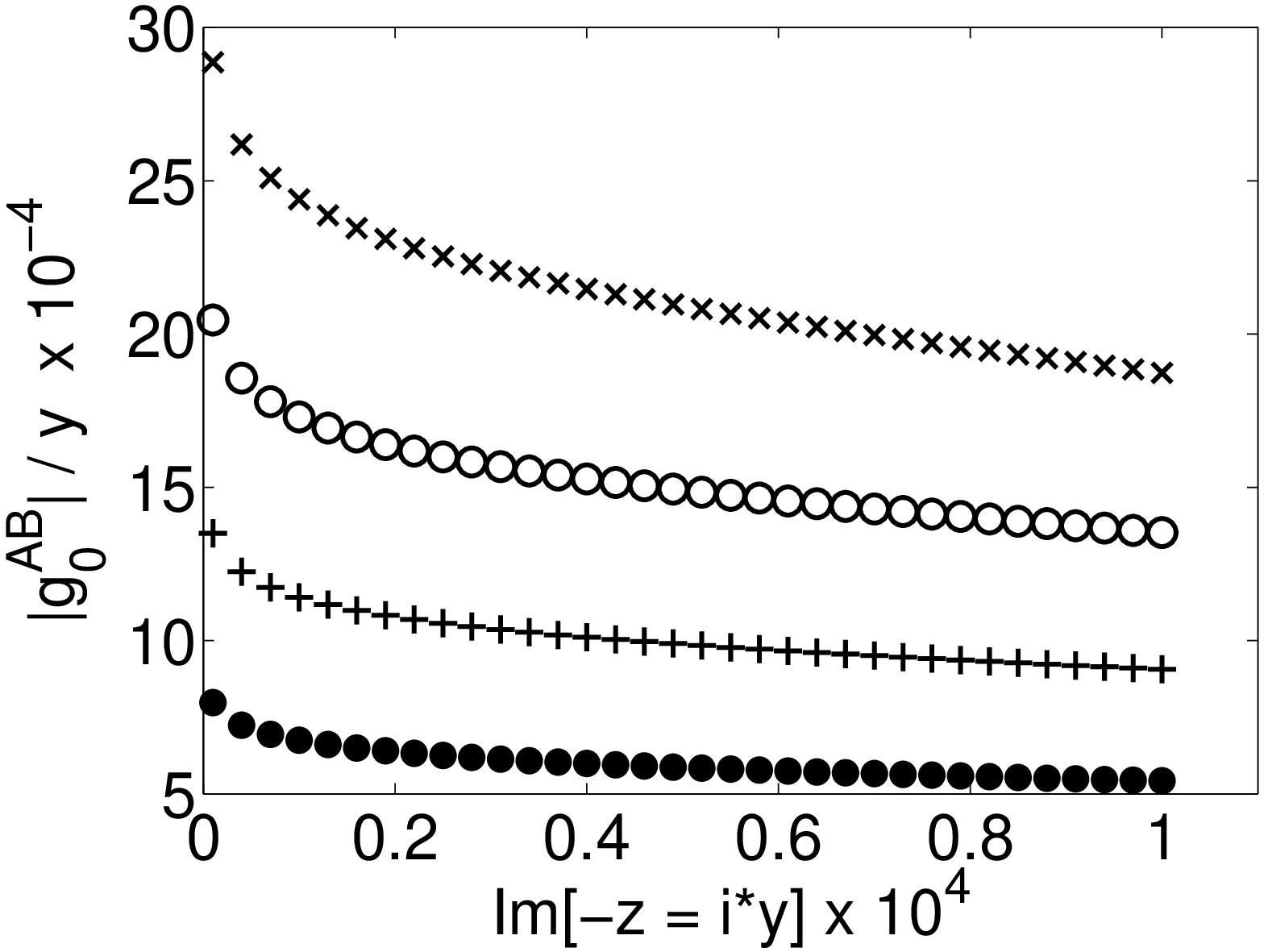,width=0.4\textwidth}
\epsfig{file=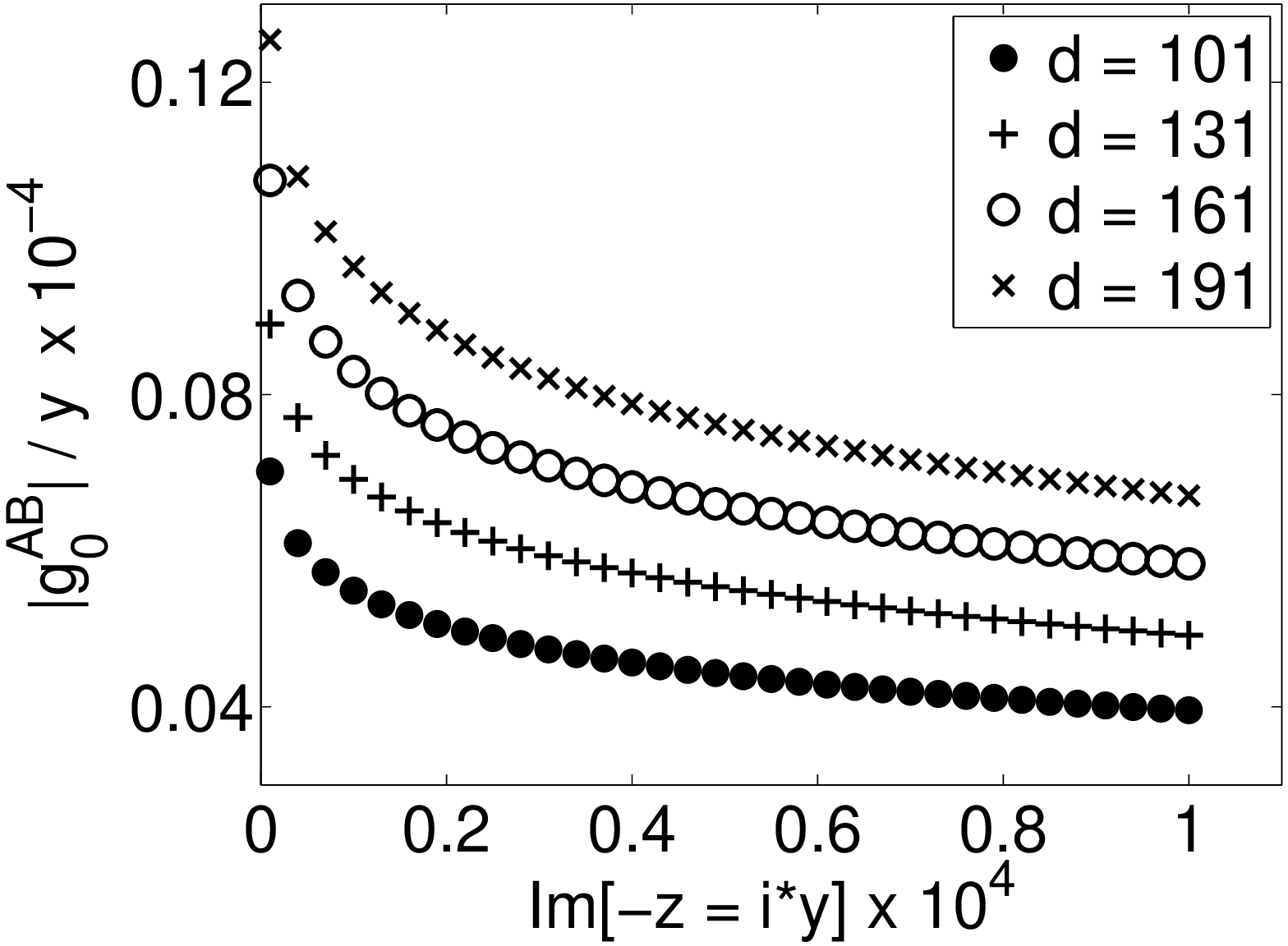,width=0.4\textwidth}
\caption{\label{gAB} Comparison of the dependence on complex energy $z$ of the Green function element $g^{AB}_0$ in a system with two vacancies on opposite sublattices. Top: in the flux free sector. Bottom: in the ground state flux sector. Vacancies have separation $d$ in the $\hat{\bf x}$ direction, with values of $d$ as indicated.
The asymptotic form in the flux free sector at small $z$ is 
$g_{0}^{AB}\propto d^{2}z\ln(-z^{2})$. 
}
\end{figure}

\begin{figure}[t]
\epsfig{file=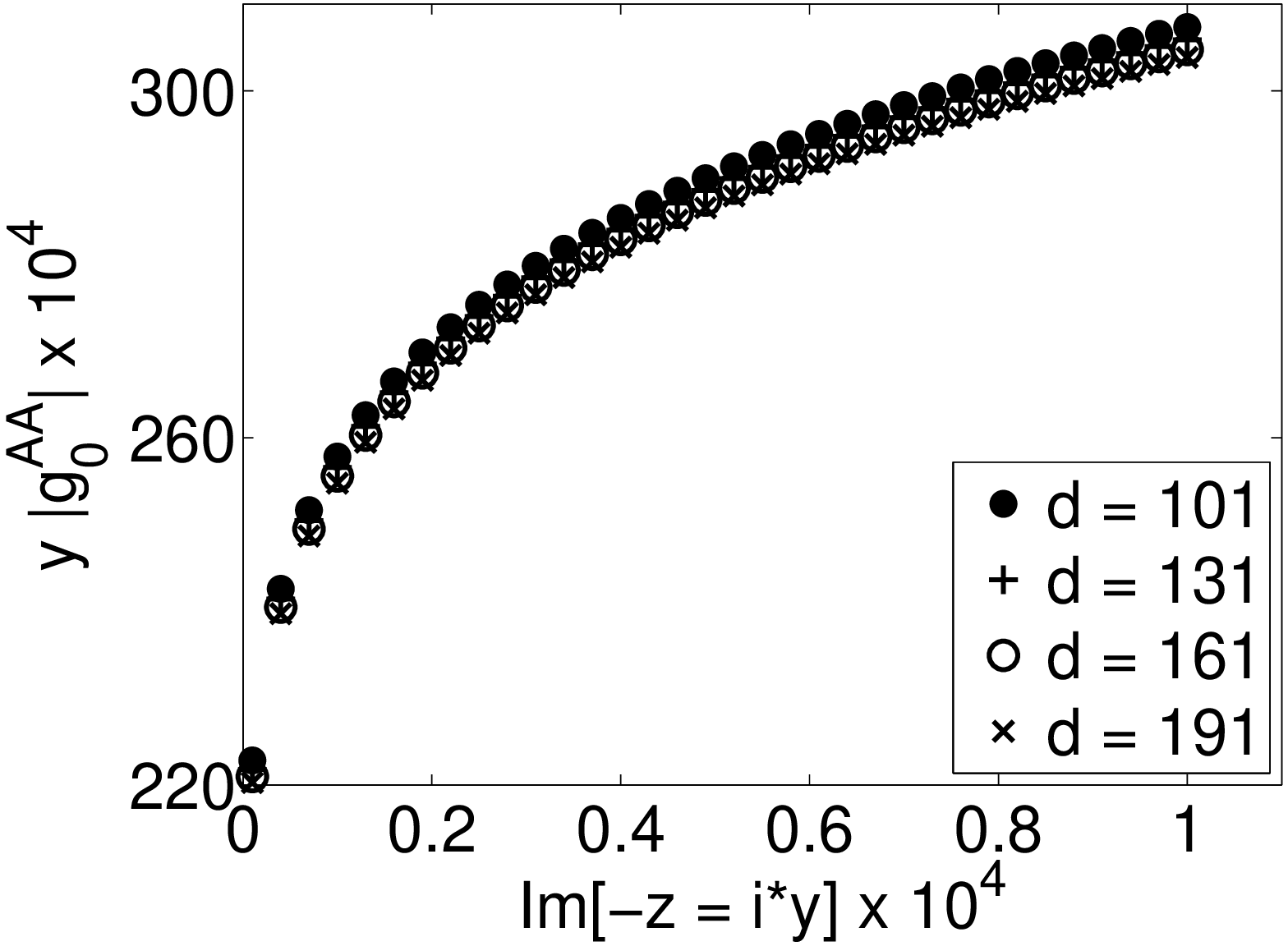,width=0.4\textwidth}
\epsfig{file=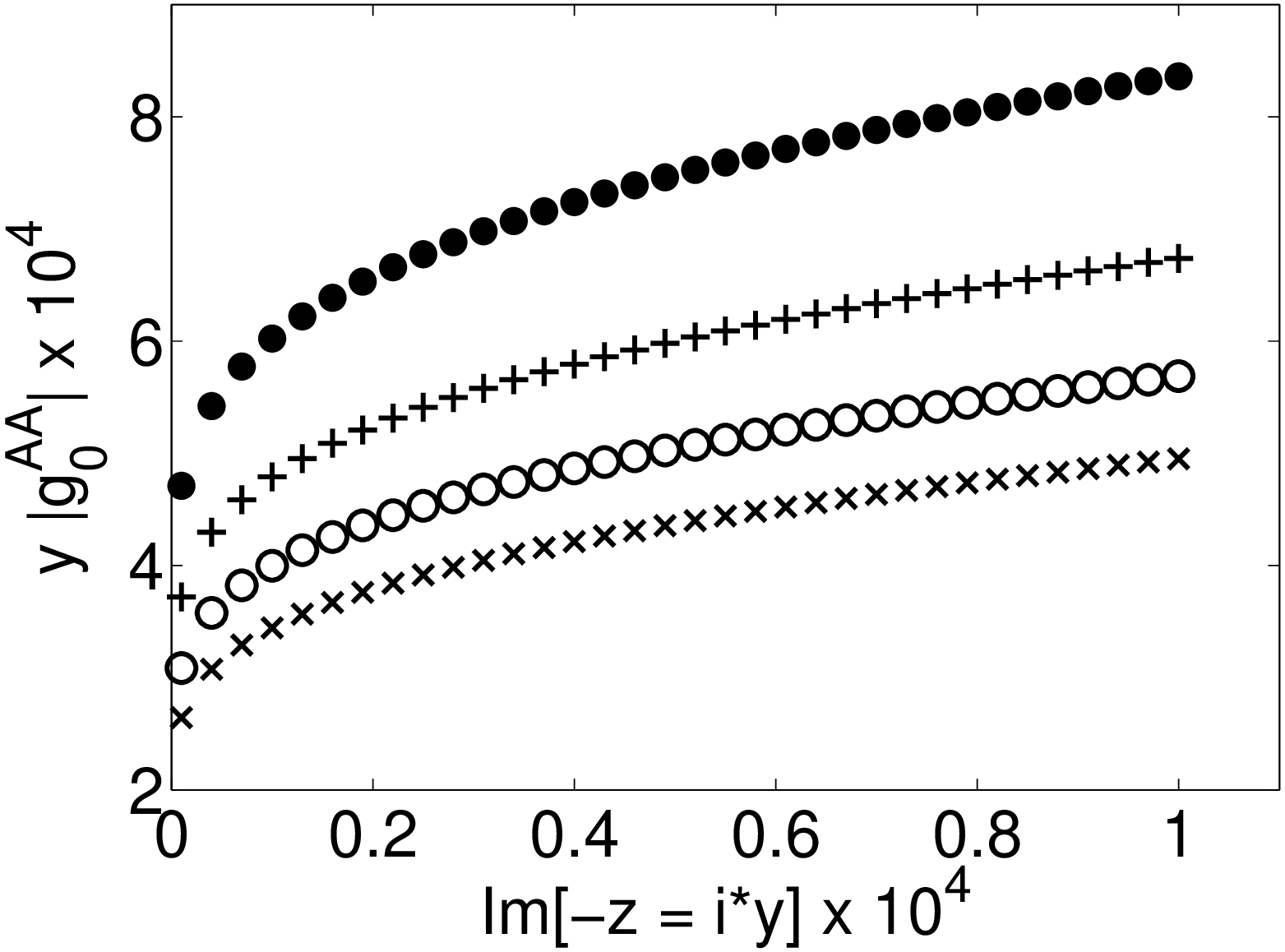,width=0.4\textwidth}
\caption{\label{gAA} As Fig.~\ref{gAB}, but for vacancies on the same sublattice. The asymptotic form for small $z$ in the flux free sector is $g_{0}^{AA}\propto [z\ln(-z^{2})]^{-1}$.}
\end{figure}

As a final step, we consider $g(z)$ in the ground state flux sector, with fluxes through each of the two vacancy plaquettes. 
Because the presence of fluxes breaks translational invariance, we no longer have analytic expressions for the
Green function of the undiluted lattice. Instead, we calculate this, and hence $g(z)$, using a numerical implementation of the $T$-matrix approach to obtain results for a pair of vacancies and fluxes, with separation in the range 100 - 200, embedded in an infinite lattice. We write
$
g_{0}^{\alpha\beta}\equiv G(0,\rr_{1},\rr_{1})
$,
with $\alpha, \beta$ labeling the vacancy sublattices. In Figs.~\ref{gAB} and \ref{gAA} we compare 
the small-$z$ behaviour of $g_{0}^{\alpha\beta}$ in the ground state and zero flux sectors, for vacancies on opposite and the same sublattices, respectively. These results demonstrate that the form of the Green function at small $z$ is indeed the same in both flux sectors, as anticipated from the idea that for low energy properties nearby fluxes effectively fuse and therefore cancel. 
The main distinction between the two flux sectors is a large scale factor relating behaviour in the two cases.

\section{Conclusions and outlook}\label{Summary}

In summary, this work builds on our observation that
the Kitaev honeycomb model 
is solvable in the presence of vacancies and the leading Zeeman coupling.
Indeed, Kitaev's original solution strategy -- to identify a non-dynamical flux field and analyse
an effective hopping problem for fermionic variables in each sector -- remains
applicable.

This is all the more remarkable as several observables change in a
fundamental way in the presence of vacancies. Firstly, the ground
state ceases to be flux free: instead, vacancies each bind a unit of the
emergent $Z_2$ flux. Secondly, the finite linear local
susceptibility of the pure system is replaced by that evidencing the
formation of a local moment around the vacancy, with an entirely free
moment whose size varies continuously with the coupling constants in
the gapped phase. Such moment formation can happen because a vacancy
locally reduces the number of constraints on the spins adjacent to it.
Thirdly, the vacancy moments interact, which is somewhat in contrast
to the ultra-short range spin correlations of the pure system. In the
gapped phase, we discover the physics of strong disorder fixed points
via a mapping to a bipartite random hopping problem. Most remarkably,
in the gapless phase, the magnetic response of two nearby vacancies on
the same sublattice is parametrically enhanced with respect to the
single-vacancy case. The resulting forms of the susceptibility are not
only a test of our assertion that vacancies bind a flux but also of
the basic description of the Kitaev model in terms of the variables
laid out in section \ref{TheModel}.

In the absence of an experimental realisation of Kitaev's honeycomb
model, it might seem premature to ask how one would go about
experimentally probing the phenomena we have described. Given the
intense interest in realising this system via a cold-atom simulator
\cite{Demler:2003}, this question is perhaps not quite so far-fetched,
particularly in view of recent developments towards single-site
microscopy in optical lattices. These may open a unique window on
introducing and locally probing the quantum state around vacancy
degrees of freedom interacting through a strongly correlated bulk. The
detection of the bound emergent $Z_2$ flux presents a particularly
exciting challenge.

In addition, with a conventional condensed matter setting in mind, our 
results illustrate for this model how disorder can serve as a probe of
quantum correlated matter. 
Local probes of magnetism, such as NMR or muon spin rotation, have a parametrically
enhanced local susceptibility around the vacancy compared to the bulk. Moreover,
since the local susceptibility of a vacancy depends on its flux state, these probes 
even give sensitivity to the flux degrees of freedom.

This behaviour is fundamentally distinct from what we found in
the case of slowly varying, weak bond disorder \cite{willans2010}. There, the
main consequence of the presence of disorder was not in the magnetic
response but rather in the heat capacity, $C\propto T^\frac{2}{1+\Delta}$, which exhibits a downward drift in its exponent with disorder strength parametrised by $\Delta \propto \langle \delta J \rangle^2$. Yet other types of disorder are comparatively featureless --
in the case of $\pm J$ bond disorder, the zero-field spectrum remains
entirely unchanged as that disorder can be `gauged away' by placing a
flux on every plaquette with a negative product of $J$'s around it.

Plenty of open issues remain. For instance, a description of the
many-vacancy behaviour based on an effective Hamiltonian would clearly
be desirable, particularly in the gapless phase, as would be a simple
formulation of the fermionic low-energy excitations of the system with
a single vacancy and its attendant flux.

Finally, there is as yet no commonly accepted classification (or,
indeed, definition) of quantum spin liquids. Whereas for the case of
gapped spin liquids, a classification via emergent gauge fields at
least seems reasonable, the situation is quite unresolved in the case
of gapless ones. Indeed, one of our result presents a step backwards:
since bond disorder leads to a {\em continuous} drift of exponents, it
is not even possible to diagnose a `Dirac spin liquid' via heat
capacity exponents without knowledge about presence and nature of random
strains. Nonetheless, the richness of phenomena induced by static
vacancies does suggest that in working towards a systematic
understanding of spin liquids, the response to disorder of a magnetic state 
of interest may present particularly valuable clues.

\section{Acknowledgements}

We thank Kedar Damle, David Huse, Dmitry Kovrizhin and Gil Rafael for valuable discussions.
The work was supported in part by EPSRC under Grant No.~EP/D050952/1.

\newpage
\appendix
\section{Projection and the physical subspace}\label{ProjectionAppend}
In this appendix we discuss the projection operator that is necessary to obtain physical states of the system from eigenstates in the enlarged Hilbert space introduced with the transformation from spins to Majorana fermions. In the extended Hilbert space of $H_{\hat{u}}$, Eq.~(\ref{quadH}), each state has at least a degeneracy of $2^{2N}$, as 
can be seen in the following manner. Define an operator  $\widetilde{D}_{j} = -i\sigma_{j}^{x}\sigma_{j}^{y}\sigma_{j}^{z} = +1$ for each lattice site in the Hilbert space of spins. In the Hilbert space of Majoranas this operator is $D_{j} = b_{j}^{x}b_{j}^{y}b_{j}^{z}c_{j}$ and commutes with the Hamiltonian $H_{\hat{u}}$ (\ref{quadH}). There are $2N$ operators $D_{j}$, each with eigenvalues $\pm 1$, which leads to the aforementioned degeneracy.

Knowing that $\widetilde{D} = +1$, it is clear that physical states of the Majorana system must have positive eigenvalue for $D_{j}$ on every site. Kitaev constructed a projection operator
\begin{equation}\label{projection}
\mathcal{P} = \dprod_{j}\left(\dfrac{1+D_{j}}{2}\right)
\end{equation}
 that ensures this.
 
\begin{figure}[t]
\epsfig{file=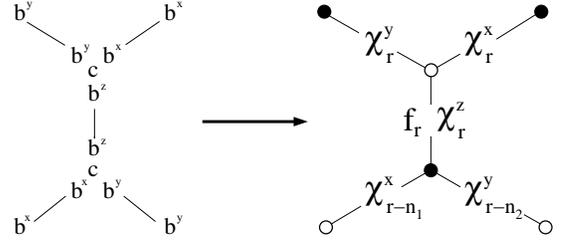,width=0.4\textwidth}
\caption{\label{refermionizationFig}The re-fermionization of Majorana fermions to complex fermions. Variables $\chi_{\rr}^{\alpha}$ are located on the bonds of the lattice and $f_{\rr}$ in the unit cells.}
\end{figure}

In the remainder of this section we review an analysis of the projection operator, expanding on Refs.~[{\onlinecite{Baskaran:2007lr,Yao-2008}}], and also show that for the observables we consider, matrix elements evaluated using eigenstates of the Hamiltonian $H_{\hat{u}}$ are the same as those obtained using the projected physical states. To further our efforts in understanding the effects of the projection operator, it is instructive to form complex fermions from the Majoranas $b^{x},b^{y},b^{z}, c$: with sites $j,k$ nearest neighbours on the $A,B$ sublattices respectively, define new variables\cite{Baskaran:2007lr,Yao-2008}
\begin{equation}\label{refermionization}
\begin{array}{l@{\qquad}l}
\chi_{\rr}^{\alpha} = \frac{1}{2}\left(b_{j}^{\alpha_{jk}}+ib_{k}^{\alpha_{jk}}\right)& \left(\chi_{\rr}^{\alpha}\right)^{\dagger} = \frac{1}{2}\left(b_{j}^{\alpha_{jk}}-ib_{k}^{\alpha_{jk}}\right)\\[2ex]
f_{\rr} = \frac{1}{2}\left(c_{A,\rr}+ic_{B,\rr}\right)&f_{\rr}^{\dagger} = \frac{1}{2}\left(c_{A,\rr}-ic_{B,\rr}\right).
\end{array}
\end{equation}
The $\chi_{\rr}^{\alpha}$ are located on the bonds of the lattice and $f_{\rr}$ in the unit cells, as shown in Fig. \ref{refermionizationFig}. In this notation, let site $j$ be in a unit cell at $\rr$. Then $\hat{u}_{jk} \equiv ib_{j}^{\alpha_{jk}}b_{k}^{\alpha_{jk}} = 2\left(\chi_{\rr}^{\alpha}\right)^{\dagger} \chi_{\rr}^{\alpha}-1$. Eigenstates of $H_{\hat{u}}$, Eq.~(\ref{quadH}), are direct products of a wavefunction  $|\chi\rangle$ for the gauge degrees of freedom, and a wavefunction $|f\rangle$ for the matter fields: $|\Phi\rangle = |\chi\rangle\otimes|f\rangle$. The choice of $|\chi\rangle$ encodes the flux sector and gauge, while $|f\rangle$ is an eigenstate of the Hamiltonian $H_{u}$, Eq.~(\ref{hopC}), which in the notation of Eq.~(\ref{refermionization}) takes the form

\begin{equation}\label{refermionizedHu}
\begin{array}{c}
M_{A} = \frac{1}{2}(M^{T}-M) \qquad M_{S} = \frac{1}{2}(M^{T}+M)\\[2ex]
H_{u} = \left(f^{\dagger}~f\right)\left(\begin{array}{rr}M_{A}&M_{S}\\-M_{S}& -M_{A}\end{array}\right)\left(\begin{array}{c}f^{\dagger}\\f\end{array}\right)\,.
\end{array}
\end{equation}

In terms of the complex fermions $\chi$ and $f$, the operators $D_{j}$ now take different forms on the $A$ and $B$ sublattices:
\begin{equation}
\begin{array}{rl}
D_{A,\rr} =& (\chi^{x~\dagger}_{\rr}+\chi_{\rr}^{x})(\chi^{y~\dagger}_{\rr}+\chi_{\rr}^{y})(\chi^{z~\dagger}_{\rr}+\chi_{\rr}^{z})(f_{\rr}^{\dagger}+f_{\rr})\,,\\[1ex]
D_{B,\rr} =& (\chi^{x~\dagger}_{\rr-\nn_{1}}-\chi_{\rr-\nn_{1}}^{x})(\chi^{y~\dagger}_{\rr-\nn_{2}}-\chi_{\rr-\nn_{2}}^{y})(\chi^{z~\dagger}_{\rr}-\chi_{\rr}^{z})\times\\[1ex]
&(f_{\rr}^{\dagger}-f_{\rr})\,.
\end{array}
\end{equation}
The projection operator can also be re-written as
\begin{equation}
\mathcal{P} = \frac{1}{2^{2N}}\left[1+\dsum_{\mu,\rr}D_{\mu,\rr}+\hspace{-3mm}\dsum_{\mu,\rr< \nu,\rr'}\hspace{-3mm}D_{\mu,\rr}D_{\nu,\rr'}+\cdots+\dprod_{\mu,\rr}D_{\mu,\rr}\right].
\label{ProjectionSum}
\end{equation}

The operator $D_{A(B),\rr}$ changes the bond fermion number on the three bonds attached to site $\rr_{A(B)}$, leaving the flux sector unchanged. It also acts on the matter state $|f\rangle$. Notice, however, that acting with $D_{j}$ on the sites at
both ends of a bond leaves the fermion number on that bond unchanged, and acting with $D_{j}$ on all sites in the lattice
leaves all bond fermion numbers unchanged.
Defining $D = \prod_{j}D_{j}$ and $\mathcal{P}'$ as the sum of all operators in $\mathcal{P}$ that change the bond fermion number in an inequivalent way, normalised by $1/2^{2N-1}$ since there are $2^{2N-1}$ terms in $\mathcal{P'}$, the projection operator can be rewritten\cite{Yao-2008} as $\mathcal{P} = \mathcal{P}'(1+D)/2$. Here $D$ gives the parity of the total fermion number: with $N_{\chi}$ and $N_{f}$ the number of bond and matter fermions respectively
\begin{equation}
D=(-1)^{N_{\chi}}(-1)^{N_{f}}\,.
\end{equation}
In this form, it is clear that the projection operator annihilates states of odd total fermion number. The complex fermion Hamiltonian $H_{u}$ in Eq. (\ref{refermionizedHu}) conserves fermion number modulo 2 and therefore commutes with $D$, which also commutes $H_{\hat{u}}$. Contrast this with $D_{j}$ that commutes with $H_{\hat{u}}$ but \textit{not} with $H_{u}$. One can then block diagonalise $H_{u}$ into blocks that act on Hilbert spaces of even and odd fermion number.

To summarise, the projection operator either annihilates a state that has odd total fermion number or transforms a state with even total fermion number to an equal weight superposition of all terms in $\mathcal{P}$ that change bond fermion number in an inequivalent way.\cite{Yao-2008}

Having established the effects of the projection operator we will now show that for a large class of operators, matrix elements evaluated using an eigenstate $|\Phi\rangle = |\chi\rangle\otimes|f\rangle$ of $H_{\hat{u}}$ are the same as those obtained using the projected physical states. 

\begin{figure}
\epsfig{file=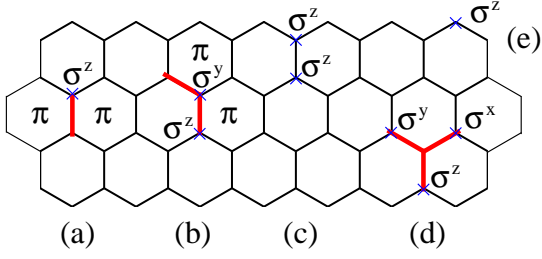,width=0.4\textwidth}
\caption{\label{operators} A graphical illustration of the final state that results from the action of various spin operators on an initial state $|\chi\rangle$ with all $\langle \hat{u}_{jk}\rangle = \langle2\left(\chi_{\rr}^{\alpha}\right)^{\dagger} \chi_{\rr}^{\alpha}-1\rangle=+1$.  We denote $u_{jk}=+1$ by black lines and $u_{jk}=-1$ by thick red lines. Spin operators act on both the gauge and matter fields but we illustrate here only the gauge fields. (a) A single spin operator $\sigma_{j}^{\alpha}$ changes the bond fermion number on a single bond and changes the flux sector, adding $\pi$ flux either side of the bond. (b) and (c) Nearest neighbour spin operators $\sigma_{j}^{\alpha}\sigma_{k}^{\beta}$ change the bond fermion number and flux sector unless they are of the form $\sigma_{j}^{\alpha_{jk}}\sigma_{k}^{\alpha_{jk}}$. (d) A three spin operator that changes bond fermion number but not the flux sector and can by undone by a term in $\mathcal{P}'$. Care must be taken with operators of this type as the terms in $\mathcal{P}'$ also act on the matter sector $|f\rangle$. (e) At the edge of the lattice, spin operators do not necessarily change the flux sector.}
\end{figure}

It is immediately apparent that if an operator $\hat{\mathcal{O}}$ changes the fermion numbers on the bonds in a manner that cannot be undone by a
term in $\mathcal{P}'$ then its expectation value is zero.
This is the case for a single spin operator or for a two spin operator that is not nearest neighbour and in the direction of the bond: if $\hat{\mathcal{O}} = \sigma_{j}^{\beta}\sigma_{k}^{\gamma}$ then $\langle \hat{\mathcal{O}}\rangle = 0$ in the ground state unless $j,k$ are nearest neighbours and $\beta=\gamma=\alpha_{jk}$. The Kitaev honeycomb model therefore has only nearest neighbour spin correlations non-zero.\cite{Baskaran:2007lr}  
The effects of various spin operators are illustrated graphically in Fig.~\ref{operators}.

For simplicity it is desirable to calculate matrix elements using an eigenstate $|\Phi\rangle = |\chi\rangle\otimes|f\rangle$. It is only permissible to do so if these matrix elements are the same as those obtained using the projected physical states. Throughout this paper we consider operators $\hat{\mathcal{O}}$ that leave  unchanged the bond fermion number. This includes the Hamiltonian and types (c) and (e) in Fig. \ref{operators}. For this class
\begin{equation}
\dfrac{\langle\Phi|\mathcal{P}\hat{\mathcal{O}}\mathcal{P}|\Phi\rangle}{\langle\Phi|\mathcal{P}\mathcal{P}|\Phi\rangle}
=\dfrac{\langle\Phi|\hat{\mathcal{O}}\mathcal{P}|\Phi\rangle}{\langle\Phi|\mathcal{P}|\Phi\rangle}
=\dfrac{\langle\Phi|\hat{\mathcal{O}}(1+D)|\Phi\rangle}{\langle\Phi|(1+D)|\Phi\rangle}
= \dfrac{\langle\Phi|\hat{\mathcal{O}}|\Phi\rangle}{\langle\Phi|\Phi\rangle}.
\end{equation}
At the steps of this derivation we have used sequentially the following facts: all spin operators commute with $\mathcal{P}$ and $\mathcal{P}^{2}=\mathcal{P}$; only the identity part of $\mathcal{P}'$ leaves the bond fermion number unchanged; and $|\Phi\rangle$ is an eigenstate of $D$. For operators of this type we are thus free to evaluate matrix elements using an unprojected eigenstate and obtain the same result as when we use the projected physical states.

\section{Green's functions of the hexagonal lattice}\label{GrFnHexLattice}
\subsection{Gapless phase}
We seek the Green function for the hexagonal lattice at small energies.  With uniform exchange coupling $J_{\alpha} = 1$, transforming Eq. (\ref{hopA}) to momentum space, the Hamiltonian and corresponding Green function are
\begin{equation}
\begin{array}{rl}
H(\qq) =& \left(\begin{array}{cc}0&f_{\qq}\\f^{*}_{\qq}&0\end{array}\right),\\
G_{0}(\qq) =& (z-H)^{-1}= \dfrac{1}{z^2-f_{\qq}f_{\qq}^{*}}\left(\begin{array}{cc}z&f_{\qq}\\f^{*}_{\qq}&z\end{array}\right),
\end{array}
\end{equation}
where $f_{\qq} = (1+e^{iq_{1}}+e^{iq_{2}})$. 
The real space Green function is then 
\begin{equation}\label{AppendixGr}
G_{0}^{\alpha\beta}(\rr) = \int G^{\alpha\beta}_{0}(\qq)e^{-i \qq.\rr}\dfrac{d^{2}\qq}{4\pi^{2}}\,.
\end{equation}

If $z$ is outside of the band then $G_{0}^{\alpha\beta}(\rr)$ can be evaluated directly for certain elements and analytically continued to the whole of the complex plane.\cite{horiguchi:1411}  Here $\rr$ denotes the unit cell and the sublattice indices $\alpha,\beta$ are given explicitly.  Using complex energy $z = E-i\epsilon$, with $E$ real and $\epsilon$ a positive infinitesimal, we define
\begin{equation}
A = \dfrac{2}{\sqrt{(z-1)^3(z+3)}}\quad B = \sqrt{\dfrac{16z}{(z-1)^3(z+3)}}\,.
\end{equation}
It can be shown that\cite{horiguchi:1411}
\begin{equation}
G^{AA}_{0}(0) = \left\{\begin{array}{c}\dfrac{z A K}{\pi}\\\\ \dfrac{-z A(K-2iK')}{\pi}\end{array}\right.\qquad \begin{array}{l}E> 1\\\\ E< 1~,\end{array}
\end{equation}
where $K=K(B^2)$ is the complete elliptic integral of the first kind, with complex parameter $k^2$ and $K'=K(1-k^2)$.  For small $z$ we find the asymptotic form given in the text.  Using the symmetry of the lattice and the defining Green functions equations, it can be shown that $zG^{AA}_{0}(0)-3G^{AB}_{0}(0) = 1$.  The Green function at an arbitrary site may be found from recursion relations involving sites closer to $\rr=0$.\cite{horiguchi:1411}

We now discuss the Green function at sites far from the origin. For $z\ll1$ and $\rr\gg 1$ the dominant contribution to the integral (\ref{AppendixGr}) is from the region of small $|f_{\qq}|$. We use the fact that the spectrum $|f_{\qq}|$ is asymptotically linear close to the Dirac nodes $\pm\QQ$, where $\QQ = (\frac{2\pi}{3},-\frac{2\pi}{3})$. Let $\qq= \pm\QQ+\pp$, then
\begin{align}
G_{0}^{AB}(\rr)
\simeq&\dfrac{\sqrt{3}}{2}\dint \left(\dfrac{-(\pp_{x}+i\pp_{y})e^{i(-\QQ.\rr-\pp.\rr)}}{z^{2}-\frac{3}{4}\pp^{2}}\right.\nonumber\\
&\left.+\dfrac{(\pp_{x}-i\pp_{y})e^{i(\QQ.\rr-\pp.\rr)}}{z^{2}-\frac{3}{4}\pp^{2}}\right)\dfrac{\sqrt{3}}{2}\dfrac{d^{2}p}{4\pi^{2}}\nonumber\\
=&-\dfrac{3i}{2}\dint \dfrac{\text{Im}\left[(\pp_{x}+i\pp_{y})e^{-i\QQ.\rr}\right]}{z^{2}-\frac{3}{4}\pp^{2}} e^{-i\pp.\rr}\dfrac{d^{2}p}{4\pi^{2}}\nonumber\\
=& \dfrac{2zi}{\sqrt{3}\pi}K_{1}[\frac{i2rz}{\sqrt{3}}]\sin\left(\QQ.\rr-\theta\right)\,,
\end{align}
where $\theta$ is the angle between $\rr$ and the x axis and $K_{1}$ is first order modified Bessel function of the second kind.
Similarly, it can be shown that
\begin{align}
G_{0}^{BA}(\rr) \simeq& \dfrac{2zi}{\sqrt{3}\pi}K_{1}[\frac{i2rz}{\sqrt{3}}]\sin\left(\QQ.\rr+\theta\right)\\
G^{AA}_{0}(\rr) \simeq&-\dfrac{2z}{\sqrt{3}\pi}K_{0}[\frac{i2rz}{\sqrt{3}}] \cos(\QQ.\rr)\,,
\end{align}
where $K_{0}$ is the zeroth order modified Bessel function of the second kind.

\subsection{Gapped phase}
We consider the parameter regime $J_{z}>J_{x}+J_{y}\geq 0$. The Green function for the gapped phase at real energies $z$ small compared to the gap $J_{z}-J_{x}-J_{y}$ can be found by a perturbative expansion in $j_{x}$ and $j_{y}$.  Let $\rr = n_{1}\nn_{1}+n_{2}\nn_{2}$ and $f_{\qq} = (J_{z}+J_{x}e^{iq_{1}}+J_{y}e^{iq_{2}})$. Then
\begin{align}
G_{0}^{AB}(\rr) =& \int \dfrac{f_{\qq}}{z^{2}-|f_{\qq}|^{2}}e^{-i \qq.\rr}\dfrac{d^{2}\qq}{4\pi^{2}}\simeq -\int\dfrac{1}{f^{*}_{\qq}}e^{-i \qq.\rr}\dfrac{d^{2}\qq}{4\pi^{2}}\nonumber\\
= &-\int J_{z}^{-1}\left(1+j_{x}e^{-i q_{1}}+j_{y}e^{-i q_{2}}\right)^{-1}e^{-i\qq.\rr} \dfrac{d^{2}q}{4\pi^{2}}\nonumber\\
=&(-1)^{|n_{1}|+|n_{2}|+1}J_{z}^{-1}~j_{x}^{|n_{1}|}~j_{y}^{|n_{2}|}{|n_{1}|+|n_{2}| \choose |n_{1}|}\nonumber\\
&\qquad n_{1}, n_{2} \leq 0,\text{ 0 otherwise.}\\
\nonumber\\
G_{0}^{BA}(\rr) \simeq&(-1)^{n_{1}+n_{2}+1}J_{z}^{-1}~j_{x}^{n_{1}}~j_{y}^{n_{2}}{n_{1}+n_{2} \choose n_{1}}\nonumber\\
&\qquad n_{1}, n_{2} \geq 0,\text{ 0 otherwise.}
\end{align}
The Green function between sites on the same sublattice can be obtained in an analogous manner. We find
\begin{widetext}
\begin{align}
\label{GAAgapped}
G_{0}^{AA}(\rr) \simeq& (-1)^{|n_{1}|+|n_{2}|+1}\dfrac{z}{J_{z}^{2}}~j_{x}^{|n_{1}|}~j_{y}^{|n_{2}|}\times\left[\dsum_{k_{1},k_{2}=0}^{\infty}{|n_{1}|+|n_{2}|+k_{1}+k_{2} \choose |n_{1}|+k_{1}}{k_{1}+k_{2} \choose k_{1}}~j_{x}^{2k_{1}}~j_{y}^{2k_{2}}\right]\nonumber\\
&\text{for sgn(n1) = sgn(n2)}\nonumber\\
G_{0}^{AA}(\rr) \simeq& (-1)^{|n_{1}|+|n_{2}|+1}\dfrac{z}{J_{z}^{2}}~j_{x}^{|n_{1}|}~j_{y}^{|n_{2}|}\times\left[\dsum_{k_{1},k_{2}=0}^{\infty}{|n_{1}|+k_{1}+k_{2} \choose |n_{1}|+k_{1}}{|n_{2}|+k_{1}+k_{2} \choose |n_{2}|+k_{1}}~j_{x}^{2k_{1}}~j_{y}^{2k_{2}}\right]\nonumber\\
&\text{for sgn(n1) $\neq$ sgn(n2)}
\end{align}
\end{widetext}

\bibstyle{apsrev.bst}
\bibliography{citeThesis}

\begin{thebibliography}{23}
\bibitem[{foo()}]{footnote:I}
\bibinfo{note}{In fact, if a vacancy is located inside the wedge of another on
  the same sublattice, it will also have nonzero components $g_{xx}$ and
  $g_{yy}$ of a size exponentially small in the inter-vacancy separation.}
\end{thebibliography}

\end{document}